\documentclass[review]{elsarticle}

\usepackage{hyperref}

\journal{Journal of \LaTeX\ Templates}

\usepackage{amsmath,amssymb,nicefrac}
\usepackage[english]{babel}
\usepackage[T1]{fontenc} 
\usepackage{color}
\usepackage{amsmath}
\usepackage{paralist}
\usepackage{amssymb}
\usepackage{amsthm,amssymb,amsmath,mathrsfs}
\usepackage{psfrag}
\usepackage{mathrsfs}
\usepackage{enumerate}
\usepackage{ae}
\usepackage[T1]{fontenc}
\usepackage{colortbl}
\usepackage{graphicx}
\usepackage{enumerate}
\newcommand{\beq}{\begin{equation}}
\newcommand{\eeq}{\end{equation}}
\usepackage{booktabs}
\usepackage{rotating}
\usepackage{wrapfig,epstopdf}
\usepackage{bm}

\newcommand{\IR}{{\rm I}\!{\rm R}}
\DeclareMathOperator{\Lin}{\mathbb{L}\mathrm{in}}

\DeclareMathOperator{\SO3}{SO(3)}
\DeclareMathOperator{\tr}{tr} %
\DeclareMathOperator{\Div}{Div} %
\def\reff#1{\mbox{\rm(\ref{#1})}}

\newcommand{\td}{\,\mathrm{d}}

\newcommand{\vect}[1]{\boldsymbol{#1}}  

\newcommand{\tens}[1]{\boldsymbol{#1}}
\def\T#1{\tens{#1}}
\def\V#1{\vect{#1}}
\def\M#1{\ensuremath{\mbox{\boldmath $\mathrm{#1}$}}}

\def\1Inv{\text{I}}
\def\2Inv{\text{II}}
\def\3Inv{\text{III}}

\newcommand{\Vhatv}{\hat{\boldsymbol \lambda}}

\newcommand{\Vhatu}{\hat{\boldsymbol u}}

\newcommand{\ra}[1]{\renewcommand{\arraystretch}{#1}}
\def\reff#1{\mbox{\rm(\ref{#1})}}

\begin{document}
\begin{frontmatter}

\title{Data-driven finite element computation of open-cell foam structures}

\author[siegen]{Tim Fabian Korzeniowski\corref{cor1}}
\author[siegen]{Kerstin Weinberg}

\cortext[cor1]{Corresponding author \\ E-mail address: tim.korzeniowski@uni-siegen.de}

\address[siegen]{Lehrstuhl f\"ur Festk\"orpermechanik, Universit\"at Siegen, D-57076 Siegen, Germany}
%

\begin{abstract}
This paper presents a model-free data-driven strategy for linear and non-linear finite element computations of open-cell foam. Employing sets of material data,
the data-driven problem is formulated as the minimization of a distance function subjected to the physical constraints
from the kinematics and the balance laws of the mechanical problem.

The material data sets of the foam are deduced here from representative microscopic material volumes. These volume elements capture the stochastic characteristics of the open-cell microstructure and the properties of the polyurethane material. Their computation provides the required stress-strain data used in the macroscopic finite element computations without postulating a specific constitutive model.
The paper shows how to efficiently derive suitable material data sets for the different (non-)linear and (an-)isotropic material behavior cases.

Exemplarily, we compare data-driven finite element computations with linearized and finite deformations 
and show that a linear kinematic is sufficiently accurate 
to capture the material's non-linearity up to 50\% of straining. 
The numerical example of a rubber sealing illustrates possible areas of application, the expenditure, and the proposed strategy's versatility.

\end{abstract}


\end{frontmatter}


\section{Introduction}
Solid foams,  typically used for thermal insulation, vibration damping  and sandwich-structured composites,
are lightweight engineering materials with a heterogeneous cellular microstructure.
Their design and computation require either strong simplifications or sophisticated material models.
To avoid these, a direct use of  measurement data in the computation would be desirable. 
Kirchdoerfer and Ortiz \cite{KirchdoerferOrtiz2016} recently introduced such a model-free computing method,  the so-called data-driven (DD) computational mechanics, which incorporates data directly into the finite element method (FEM), thus bypassing the need for an explicit  constitutive model.
The numerical approach was supplemented by mathematical investigations in Conti et al. \cite{ContiOrtiz2017}, demonstrating that the DD-FEM 
comprises the classical definition of solid mechanics boundary value problems (BVPs).
Whereas the original work \cite{KirchdoerferOrtiz2016} was proposed in the framework of linear elasticity,
several extensions to two- and three-dimensional problems followed \cite{kirnoise,kirdynamics,leygue2017data,KorzeniowskiWeinberg2019}.
Further contributions focus on
efficient data management strategies \cite{korzeniowski2021multi,eggersmann2021model,eggersmann2021efficient},
and extend the method to
elastodynamics \cite{kirdynamics,nguyen2020variational},
non-linear elasticity \cite{nguyen2018data,ibanez2018manifold,platzer2020finite},
inelasticity \cite{reese,ibanez2019hybrid},
and diffusion \cite{korzeniowski_diffPAMM}. All these contributions consider the required data to be known.

For solid mechanic problems, the material data set which replaces the constitutive model needs to represent strains and stresses in the dimension of the problem. They have to encompass an almost infinite number of data tuples, covering every possible mechanical state under consideration. Such data sets can be gained experimentally, using in-situ computed tomography for example, but these sorts of measurements are expensive and for many problems not available. An alternative way of data acquisition can be the use of computational  `material testing' in the sense of sample calculations for the heterogeneous material. In the classical approach, representative volume elements (RVEs), which allow for a detailed description of the microscopic   characteristics, are used to derive   effective material properties \cite{allaire1992homogenization}. 
For the finite element analysis (FEA) of engineering components  this technique leads directly to  computational homogenization and FE$^2$ computations \cite{miehe1999computational,feyel2000fe2}, where the two- or multi-scale problem is decoupled into nested subproblems,
i.e. a microscopic BVP and a macroscopic BVP. 
For the  macroscopic BVP to be solved, the required effective quantities are obtained from the solutions of the many microscopic BVPs at the  material points of the discretized continuum, \cite{abdulle2012heterogeneous,cioranescu2008periodic}.

For foam-like materials, homogenization  strategies are chosen in \cite{kanaun2007representative,marvi2018modelling} for example.
Here, we  focus  specifically on an open-cell elastomeric foam. The mechanical properties of such foams are determined by the cellular microstructure and by the material's properties.  In our study, the foam is made of polyurethane (PUR),  a common elastomer produced by a reaction of an isocyanate with a polyol in the presence of a catalyst and some additives. Adding a certain amount of water to the reaction produces $\mathrm{CO_2}$, which results in the formation of gaseous bubbles, \cite{kaiser2015kunststoffchemie,bayer1947di}. After solidification, a foam with defined {porosity} but an irregular cellular microstructure is `baked'. This variability can be captured by stochastically generated RVEs, based on the same statistical parameters but with different expressions of the characteristics from one realization to another.

The material data sets were generated synthetically in most previous publications on DD-FEM \cite{KirchdoerferOrtiz2016,nguyen2018data,reese}.
In our contribution, we suggest utilizing the stochastic RVEs to generate the material database required for data-driven computations either a priori or on the fly. Exemplarily, we consider data tuples of a  deformation or  strain measure $\T e$
and a corresponding stress state $\T s$, which give at datum $i$ the tuple $\V z_i=(\T e,\T s)_i $. All tuples may depend on space and time; we drop the arguments here for readability. Also, in view of finite deformation mechanics,  we leave the specific strain and stress measures open at this point. Then, a typical material data set $\mathcal{D}$ consists of a finite number of local states
\begin{align*}
   \mathcal{D}  = \{\V z_i, i=1,\dots,n \}\, .
\end{align*}
The data-driven solution is then the field of states $\M z\in \mathcal{D}_g=\mathcal{D}\times\mathcal{D}\times\mathcal{D}\times\dots$ of the global state space $\mathcal{D}_g$  that is closest to satisfying kinematic compatibility and physical balance laws.  Denoting the set of compatible deformation  fields in equilibrium by $\mathcal{C}$, we may formulate the data-driven problem as
\begin{align}\label{eq:problemS}
  \mathcal{S}=\text{argmin} \left\{d(\M z,\mathcal{C}) \,|\, \M z \in \mathcal{D}_g  \right\}\,
\end{align}
where    $\mathcal{S} $ is the (approximate) solution set corresponding to the total data set $\mathcal{D}_g$ and $d(\cdot,\cdot) $ is an appropriate metric measuring the  distance
 between the assigned data and the compatible solution.  This distance is   typically measured in an energy norm.

In order to provide the  data set $\mathcal{D}$, the stochastic RVE representing a specific foam must be 
subjected to all sorts of deformation states. From the solution of these  microscopic BVPs,  the corresponding stress states are obtained. In such a way we generate the state space tuples  $\V z_{i}$ that  describes the macroscopic  behavior of the heterogeneous material.

The paper is organized as follows: In Section~\ref{sec:basics} we outline the concept of the DD-FEM for a general, non-linear kinematic.  The corresponding finite element approximation is also described here for two different tuples of strain and stress measures. In Section~\ref{sec:rve} we focus on the generation of the data set. We introduce open-cell foam volumes whose microstructure is generated stochastically. Briefly, we describe the RVE model  before we discuss strategies to generate the data set $\mathcal{D}$ for different cases of linear or non-linear and isotropic or anisotropic material behavior. In the following Section \ref{sec:numeric}, we present three numerical examples for data-driven computations. At first, we compare  the different DD-FEM strategies for a simple rod under tension in Section~\ref{sec:stab}.
In \ref{sec:exa2} and \ref{sec:exa3} two and three-dimensional computations  of a rubber sealing illustrate the pros and cons of the method for engineering applications. The paper closes with a  summary in Section~\ref{sec:summary}.

\section{Governing equations}\label{sec:basics}
We  consider a solid of domain $\Omega_0$ in its reference configuration and deforming to 
the current configuration under the action of external body forces $\rho_0\V B$ and boundary tractions $\V T$; the fields in capitals refer to the reference configuration.  The solid's  deformation $\V \varphi (\T X) : \Omega_0  \mapsto \IR^3 $  is completely described by the deformation gradient
\begin{align}\label{eq:defo}
    \T F = \nabla {\vect{\varphi}}(\V X)\,.
\end{align}
The  work conjugate stress is the first Piola-Kirchhoff tensor $\T P$;  the second Piola-Kirchhoff tensor follows via $\T S = \tens{F}^{-1}\tens{P} $.
The stresses fulfill the linear and the angular momentum balances
\begin{align}\label{eq:GGWinP}
\Div \T P + \rho_0 \vect{{B}} &= 0 \qquad\quad \mbox{ in } \Omega_0
\\\label{eq:MomentenGGWinP}
    \tens{P}^T\tens{F}-\tens{F}\tens{P}^T  &=0
\end{align}
and consequently is $\T S=\T S^T$. The solid is subjected to geometrical and static boundary conditions at its boundaries $\Gamma_0$ and $\Gamma_1$ with outward unit normal $\V  N$,
\begin{align}
\V \varphi &= \bar{\V \varphi} \qquad\quad \mbox{on }\Gamma_0 \,, \\\label{BCstatic}
\tens{P} \,\V  N &= {\vect{T}} \qquad\quad \mbox{on }\Gamma_1 \,.
\end{align}
where $\Gamma_0 \cup \Gamma_1 =\partial\Omega_0$ and $\Gamma_0\cap\Gamma_1=\emptyset$.
We aim at solving this BVP in its data-driven form \reff{eq:problemS}.
To do so, we first seek a proper metric of the phase space, using hyperelastic strain-energy densities as a guide that typically have terms of the form $|\T F|^p$ and $|\T F^q|$. 

\subsection{Data-driven problem in \textbf{F}, \textbf{P} }\label{sec:FB}
Let us start with the simplest case, namely a Neo-Hooke-like model with $p=2$ and an elastic strain-energy density as a function of the deformation gradient, 
\begin{align}\label{EnergiedichtePsiF}
    \Psi^e (\tens{F}) =   \frac{\mu }{2}     \left( \T F :\T F - 3\right) \,.
\end{align}
Via a Legendre transformation, the complementary energy density as a function of the stress tensor $\T P$
\begin{align*}
  \Psi' (\tens{P}) &= \max_{\T F \in \Lin}\left[ \T P:\T F - \frac{\mu }{2}     \left( \tens{F} : \tens{F}  - 3 \right)  \right]\\
&=  \frac{1}{2\mu }      \tens{P} : \tens{P}  + \frac{3\mu }{2}\,.
\end{align*}
Thus, the distance {$d(\cdot ,\cdot) $} between the compatible solution and the assigned data  can be measured locally with a  penalty functional of the form
\begin{align}\label{lokaleStraffunktionPF}
\bar{\Psi} (\tens{F},\tens{P}) &=\min_{(\tens{F}',\tens{P}')\in \mathcal{D}} \left(\Psi^e(\tens{F}-\tens{F}') + \Psi' (\tens{P}-\tens{P}')\right)
\\\nonumber
                               &=\min_{(\tens{F}',\tens{P}')\in \mathcal{D}} \left(\frac{\mu_0}{2}  (\tens{F}-\tens{F'} ) : (\tens{F}-\tens{F'})
                                                                                  +\frac{1}{2\mu_0}  (\tens{P}-\tens{P'} ) : (\tens{P}-\tens{P'})  \right)
\end{align}
where the data set $\mathcal{D}=\{ (\tens{F},\tens{P})_1, \dots, (\tens{F},\tens{P})_n\}$  contains tuples of values of the deformation gradient $\T F$ and the first Piola-Kirchhoff stress tensor $\T P$.  We presume them to be physical meaningful and to fulfill eq.~\reff{eq:MomentenGGWinP}.
The  parameter $\mu$ calibrates  units and magnitudes and is purely numerical in nature; we therefore write $\mu_0$ in the following. When minimizing the local penalty functional \reff{lokaleStraffunktionPF}, the kinematics of the deformation \reff{eq:defo} as well as the momentum balances \reff{eq:GGWinP} and \reff{eq:MomentenGGWinP} must hold for all admissible states $\tens{F}$ and $\tens{P}$.
Equivalently, 
we can formulate a  global penalty functional for the whole domain or a part of it
\begin{align*}
\mathcal{W}^\text{pen}=\int_{\Omega_e} \bar{\Psi} (\tens{F},\tens{P}) \td \Omega \qquad \rightarrow \text{min.},
\end{align*}
and  we may likewise demand the equilibrium conditions in global form.

We want to start with the local form \reff{lokaleStraffunktionPF}. The data-driven BVP then reads:  $\bar{\Psi}^*  \rightarrow \text{min.}$ with functional
\begin{align*} 
\bar{\Psi}^* =
\min_{(\tens{F}',\tens{P}')\in \mathcal{D}} &\left(\Psi^e(\tens{F}-\tens{F}') + \Psi' (\tens{P}-\tens{P}')\right)\\
 &+ \vect{\lambda}\left( \nabla\cdot \tens{P} + \varrho_0\V B\right) + \vect{\lambda}_2 : \left(\tens{F}\tens{P}^T -\tens{P}^T \tens{F}  \right)
 \,.
\end{align*}
The additional Lagrange multiplier field  $\vect{\lambda}$ enforces the static equilibrium \reff{eq:GGWinP} and is a displacement-like vector field,
the tensor-valued field $\vect{\lambda}_2$ enforces the angular momentum balance \reff{eq:MomentenGGWinP}.
For  the optimization of $\bar{\Psi}^*$ with a finite element  discretization in Lagrangian form, we formulate the deformation gradient with the local displacements  $\V u$
\begin{align}\label{defo:Igradu}
   \tens{F} = {\tens{I} + \nabla \V u}
\end{align}
so that the functional depends on $\V{u}$, $\tens{P},\vect{\lambda}$ and $ \vect{\lambda}_2$.
Assuming additionally that the optimal data tuples  $(\tens{F}^*,\tens{P}^*)$ are already determined,   the functional to be optimized reads
\begin{align*}
\bar{\Psi}^*(\V{u},\tens{P},\vect{\lambda}, \vect{\lambda}_2)  =&
\frac{\mu_0}{2}  (\tens{I} + \nabla \V u-\tens{F}^* ) : (\tens{I} + \nabla \V u-\tens{F}^*)+\frac{1}{2\mu_0}  (\tens{P}-\tens{P}^* ) : (\tens{P}-\tens{P}^*)
\\ 
 &+ \vect{\lambda}\cdot (\nabla\cdot{\tens{P}} +  \varrho_0\V B )
  + \vect{\lambda}_2 : \left((\tens{I} + \nabla \V u)\tens{P}^T -\tens{P}^T (\tens{I} + \nabla \V u)  \right)
\end{align*}
and its total variation gives the equations:
\begin{equation}
\begin{aligned}\label{eq:variPF}
\delta_{\V u}\bar{\Psi}^* =0: \quad \rightarrow\quad&&
\mu_0{\nabla \delta\V u} :({\tens{I} + \nabla \V u}-\T F^*)=&\T \lambda_2:(\T P^T {\nabla \delta\V u} - {\nabla \delta\V u}\, \T P^T )
\\ 
\delta_{\T P}\bar{\Psi}^*=0: \quad \rightarrow\quad&&
\tfrac{1}{\mu_0}\delta\T P:(\T P-\T P^*)+\vect{\lambda}\cdot(\nabla\cdot\delta\T P)
=&\T \lambda_2:(   \delta\T P^T \, \nabla \V u + \nabla \V u \;\delta\T P^T )
\\ 
\delta_{\T \lambda}\bar{\Psi}^*=0: \quad \rightarrow\quad&&
  \delta\vect{\lambda} \cdot\left(\nabla\cdot \T P+\varrho_0 \V B\right) =&0
\\
\delta_{\T \lambda_2}\bar{\Psi}^*=0: \quad \rightarrow\quad&&
   \delta\vect{\lambda}_2 :\left( \nabla \V u\tens{P}^T -\tens{P}^T  \nabla \V u  \right)=&0
\end{aligned}
\end{equation}

We want to express these equations  in terms of the unknown displacements $\vect{u}$ and the 
the multiplier field $\vect{\lambda}$, which   can conveniently be approximated  by finite elements. We start with eq.~\reff{eq:variPF}$_4$,  and note that for any $\delta\vect{\lambda}_2$ its total variation equals the sum of the right-hand sides of eq.~\reff{eq:variPF}$_1$ and \reff{eq:variPF}$_2$. Consequently, the sums of the left-hand sides 
also equal.  Both terms are now formulated in the integral mean over an (element) domain $\Omega_e$,
and the second term is re-arranged using Gauss's theorem and $\delta\T P=0$ at the boundary. With eq.~\reff{defo:Igradu} we obtain
\begin{align}\label{eq:vari1gleich2}
    \int_{\Omega_e} \mu_0{\nabla \delta\V u} :(\T F-\T F^*)
    \td \Omega
=-    \int_{\Omega_e} \frac{1}{\mu_0}\delta\T P :(\T P-\T P^*+\mu_0\nabla\vect{\lambda})  \td \Omega \,.
\end{align}
At next we integrate eq.~\reff{eq:variPF}$_3$  over domain $\Omega_e$ and obtain after re-arrangement with the boundary conditions \reff{BCstatic}
\begin{align}
\label{eq:variPF32}
    \int_{\Omega_e} \nabla \delta\vect{\lambda} : \T P \td \Omega &= \int_{\partial\Omega_e} \delta\vect{\lambda} \cdot \V T \td \Omega + \int_{\Omega_e} \delta\vect{\lambda} \cdot \varrho_0 \V B \td \Omega =:\V f^\text{ext} 
    \,.
\end{align}
For further evaluation of eq.~\reff{eq:variPF32} we need an expression for the stresses $\T P$,  which we want to deduce from relation~\reff{eq:vari1gleich2}. We remark that a classical finite element  approximation provides the displacements as the primary solution. 
With the argument, that then the left-hand side of \reff{eq:vari1gleich2} is close to zero, the  bracketed term on the right-hand side of \reff{eq:vari1gleich2} must vanish for all $\delta \T P$.  Thus, we derive for the stresses the expression
\begin{align}\label{eq:SpannungP}
    \T P=\T P^*-\mu_0\nabla\vect{\lambda}
\end{align}
which we now substitute into  \reff{eq:variPF}$_3$  resp. \reff{eq:variPF32}. What remains is
\begin{align}\label{eq:vari33}
    \int_{\Omega_e} \mu_0 \nabla \delta\vect{\lambda} : \nabla \vect{\lambda}  \td \Omega = \V f^\text{ext} -
    \int_{\Omega_e}  \nabla \delta\vect{\lambda} : \T P^*   \td \Omega
\end{align}
where $ \V f_e$ abbreviates the right-hand side of \reff{eq:variPF32}. Finally, we also state the integral form of eq.~\reff{eq:variPF}$_1$,
\begin{align}\label{eq:vari12}
    \int_{\Omega_e} \mu_0 \nabla \delta\vect{u} : \nabla \vect{u}  \td \Omega &=
    \int_{\Omega_e} \mu_0 \nabla \delta\vect{u} : (\T F^* -\T I)   \td \Omega \,.
\end{align}
Equations \reff{eq:vari33} and \reff{eq:vari12} formulate the data-driven problem we are looking for. Obviously, the resulting system of equations is coupled but linear in $\V u$ and $\V \lambda$. The reason for this linearity is the stress expression \reff{eq:SpannungP}. Without this approximation it is not possible to reduce the problem to the unknown fields $\V u$ and $\V \lambda$ for the energy density \reff{EnergiedichtePsiF}.

\subsection{Finite element formulation in \textbf{F}, \textbf{P} }
For the numerical simulation we decompose the domain $\Omega_0$  of dimension $\mathrm{d}$ into finite elements   $\Omega_e$,
\begin{align*}
    \Omega_0 \approx \tilde{\Omega}_0 = \bigcup_{\mathcal{E}} \Omega_e
\end{align*}
where ${\mathcal{E}} =\{1,\dots, n_{el}\}$ is the set of all elements and $\cup$ abbreviates the assembly. We employ the usual ansatz for the unknown fields and their variations
\begin{equation}\label{femansatzMatrix}
    \begin{aligned}
        \V u(\V x) \approx \tilde{\V u}(\V x) &=\M{N}  \Vhatu  \qquad\qquad
        \delta \tilde{\V u}(\V x)  =\M{N}\, \delta\Vhatu
        \\
            \V \lambda(\V x) \approx \tilde{\V \lambda}(\V x) &=\M{N}  \Vhatv  \qquad\qquad
        \delta \tilde{\V u}(\V x)  =\M{N}\, \delta\Vhatv 
    \end{aligned}
\end{equation}
where the matrix $\M{N}$ contains the shape functions $N_k$ for every degree of freedom  $k=1,\dots,n_\text{dof}$. The vectors $\Vhatu$  and $\Vhatv $ denote the nodal displacements
and Langrange parameters. The gradients are summarized in matrix $\M{B}\equiv \nabla \M{N}$;  the subscript $e$ refers to one element.

In that way we obtain the discretization of eqs.~\reff{eq:vari33} and \reff{eq:vari12}
\begin{equation}\label{DD:statik3D:FEMgleichungPF}
    \begin{aligned} 
       \int_{\Omega_e}\mu_0 {\M{B}^e}^T {\M{B}^e}   \td \Omega \  \Vhatu_e
        &=
       \int_{\Omega_e} \mu_0 {\M{B}^e}  (\M F^* - \M I) \td \Omega
       \\ 
      \int_{\Omega_e}\mu_0 {\M{B}^e}^T  {\M{B}^e}     \td \Omega  \ \Vhatv_e   &=  \M{f}^e
      -\int_{\Omega_e}  {\M{B}^e}^T{\M{P}^*}    \td \Omega
    \end{aligned}
\end{equation}
with the data tuples $(\M F^*,\M P^*)$ in vector form and the elementwise force vector
\begin{align}\label{eq:Kraftvektor}
    \M{f}^e =     \int_{\Omega_e}  \varrho_0\M{N}^T\M{B} \td \Omega + \int_{\Gamma_{1e}} \M{N}^T\M{T}\td \Gamma \,.
\end{align}
The resulting finite element system is given in Appendix~1.

\subsection{Data-driven problem in \textbf{C}, \textbf{S} }\label{sec:DD_CS}
To indeed account for a non-linear kinematic of the deformation, an energy density with a higher order term is necessary. We choose $q=4$, i.e. $|\T F^4|$ in density \reff{EnergiedichtePsiF} and describe the deformation with the invariant right Cauchy-Green tensor $\T C=\T F^T F$. The elastic strain-energy density is then 
\begin{align*}
    \Psi^e (\tens{C}) = \frac{\mu_0}{2}     \left( \tens{C} : \tens{C}   -3\right) \,.
\end{align*}
with the corresponding conjugate, the second Piola-Kirchhoff stress tensor $\tens{S}$.
Again, via a Legendre transformation we obtain the complementary energy density,
\begin{align*}
  \Psi' (\tens{S}) &=  \frac{1}{2\mu_0}      \tens{S} : \tens{S}  + \frac{3\mu_0}{2} \,,
  \end{align*}
and so the local penalty functional  reads
\begin{align}\label{lokaleStraffunktionSF}
\bar{\Psi} (\tens{C},\tens{S}) 
                               &=\min_{(\tens{C}',\tens{S}')\in \mathcal{D}} \left(\frac{\mu_0}{2}  (\tens{C}-\tens{C'} ) : (\tens{C}-\tens{C'})
                                                                                  +\frac{1}{2\mu_0}  (\tens{S}-\tens{S'} ) : (\tens{S}-\tens{S'})  \right)
\end{align}
where the data set now contains $(\tens{C},\tens{S})$ tuples,
$\mathcal{D}=\{ (\tens{C},\tens{S})_1, \dots, (\tens{C},\tens{S})_n\}$. The minimization of functional~\reff{lokaleStraffunktionSF} has to be performed under the equilibrium constraint \reff{eq:GGWinP}, i.e. $\Div (\tens{F}\tens{S})  + \varrho_0\V B =0$,
the balance of angular momentum is fulfilled a priori.
Expressing $\tens{C} $  with the displacement field $\V u$
\begin{align*}
  \tens{C} 
            = {{\nabla \V u} +  {\nabla \V u}^T + {\nabla \V u}^T {\nabla \V u} + \tens{I}}
\end{align*}
we obtain the functional to be optimized in terms of  $\V{u}$, $\tens{S},\vect{\lambda}$. For optimal data tuples $(\tens{C}^*,\tens{S}^*)$ it reads
\begin{align*}
\bar{\Psi}^*(\V{u},\tens{S},\vect{\lambda})  =&
\frac{\mu_0}{2}  ({{\nabla \V u} +  {\nabla \V u}^T + {\nabla \V u}^T {\nabla \V u} + \tens{I}}-\tens{C}^* )^2 : \tens{I}
+\frac{1}{2\mu_0}  (\tens{S}-\tens{S}^* )^2:\tens{I}
\\ 
 &\qquad + \vect{\lambda}\cdot (\nabla\cdot{({\tens{I} + \nabla \vect{u}})\tens{S}} +  \varrho_0\V B )
\end{align*}
and gives the variational equations:
\begin{equation}
\begin{aligned}\label{eq:variSC}
\delta_{\V u}\bar{\Psi}^* =0:\quad  \rightarrow \quad&&
2\mu_0\left({{\nabla \delta \V u} +  {\nabla \delta \V u}^T {\nabla \V u}} \right):(\nabla \V u +  {\nabla \V u}^T + {\nabla \V u}^T {\nabla \V u} &+ \tens{I} -\T C^*)  \\
\phantom{\delta_{\V u}\bar{\Psi}^* =0:\quad  \rightarrow \quad}&& + \vect{\lambda}\cdot \left(\nabla\cdot{\left( \nabla \delta \vect{u}\,\tens{S}\right)} \right)  &=0 \qquad
\\ 
\delta_{\T S}\bar{\Psi}^*=0: \quad \rightarrow\quad&&
\tfrac{1}{\mu_0}\delta\T S:(\T S-\T S^*)+\vect{\lambda}\cdot \left(\nabla\cdot{\left(({\tens{I} + \nabla \vect{u}})\delta \tens{S}\right)} \right) &=0
\\ 
\delta_{\T \lambda}\bar{\Psi}^*=0: \quad \rightarrow\quad&&
  \delta\vect{\lambda} \cdot\left(\nabla\cdot ({\tens{I} + \nabla \vect{u}})\tens{S} +\varrho_0 \V B \right) &=0
\end{aligned}
\end{equation}
Again, we want to express these equations with the unknown displacements $\vect{u}$ and the multiplier field $\vect{\lambda}$ in an integral mean over a domain $\Omega_e$. From eq.~\reff{eq:variSC}$_1$ we obtain after re-arrangement
\begin{align}\label{eq:variSC12}
2\mu_0 \int_{\Omega_e}\left({{\nabla \delta \V u} +  {\nabla \delta \V u}^T {\nabla \V u}}\right)&:({{\nabla \V u} +  {\nabla \V u}^T + {\nabla \V u}^T {\nabla \V u} + \tens{I}} -\T C^*)
    \td \Omega   \\\nonumber \hfill 
& = \int_{\Omega_e}\nabla \vect{\lambda} : \nabla \delta\vect{u}\, \tens{S}    \td \Omega \,.
\end{align}
The eq.~\reff{eq:variSC}$_3$  becomes
\begin{align}
    \label{eq:variSC32}
    \int_{\Omega_e} \nabla \delta\vect{\lambda}\, {\left(\tens{I} + \nabla \vect{u}\right)^T} : \tens{S} \td \Omega = \int_{\Omega_e} \delta\vect{\lambda} \cdot \V T \td \Omega + \int_{\Omega_e} \delta\vect{\lambda} \cdot \varrho_0 \V B \td \Omega &=: \V f^\text{ext}
    \,,
\end{align}
and from eq.~\reff{eq:variSC}$_2$ it follows
\begin{align*}
    \int_{\Omega_e} \frac{1}{\mu_0} (\T S-\T S^*-\mu_0\nabla\vect{\lambda} {\left(\tens{I} + \nabla \vect{u}\right)^T}):\delta\T S   \td \Omega &= 0 \,
\end{align*}
from where we derive the expression for the stresses,
\begin{align*}
    \tens{S}=\tens{S}^*+\mu_0\nabla\vect{\lambda} {\left(\tens{I} + \nabla \vect{u}\right)^T}
\end{align*}
which now depend on the displacements and the multiplier field and are clearly non-linear. This expression is put into eq.~\reff{eq:variSC}$_3$ resp. \reff{eq:variSC32}, and it results
\begin{align}\label{eq:variSC33}
    \int_{\Omega_e} \nabla \delta\vect{\lambda} {\left(\tens{I} + \nabla \vect{u}\right)^T} : \left(\tens{S}^*+\mu_0\nabla\vect{\lambda} {\left(\tens{I} + \nabla \vect{u}\right)^T}\right) \td \Omega &= \V f^\text{ext}
\end{align}
where $\V f^\text{ext}$ again abbreviates the right side of \reff{eq:variSC32}. The eqs.~\reff{eq:variSC12} and \reff{eq:variSC33}  form the wanted coupled and non-linear system of equations for the fields $\vect{u}$ and $\vect{\lambda}$.  We summarize them again as residual equations, writing for the sake of clarity~$\tens{F}(\vect{u})={\tens{I} + \nabla \vect{u}}$:
\begin{align}
    \label{eq:variSCsystemR1}
    \int_{\Omega_e} {\nabla \delta \V u}^T    : \bigg(  2\mu_0  {\tens{F}(\vect{u})^T}\Big(\T F(\vect{u})^T \T F(\vect{u}) -\T C^*\Big)&  \\\nonumber
  -\nabla \vect{\lambda}^T \tens{S}^*    -    \mu_0  \nabla \vect{\lambda}^T \nabla\vect{\lambda} {\tens{F}(\vect{u})^T}   \bigg)     \td \Omega&=0
    \\\label{eq:variSCsystemR2}
    \int\limits_{\Omega_e}    \nabla \delta\vect{\lambda} : \left(    \mu_0{\tens{F}(\vect{u})}  \nabla\vect{\lambda} {\tens{F}(\vect{u})^T}   +    {\tens{F}(\vect{u})}  \tens{S}^*    \right) \td \Omega    - \V f^\text{ext}& = 0 \,.
\end{align}

\subsection{Finite element formulation in \textbf{C}, \textbf{S} }
The finite element  discretization of the system (\ref{eq:variSCsystemR1}-\ref{eq:variSCsystemR2}) is  performed with the ansatz~\reff{femansatzMatrix}. The deformation gradient is calculated in each integration point  as
\begin{align*}
    \M{F}^e = \M I +  \M{B}^{e}  \Vhatu^e
\end{align*}
where $\M{F}^e$ has the suitable vector form. The discretized system of residual equations is then:
\begin{equation}
    \label{eq:variSCsystemRfem}
\begin{aligned}
    {\M R}_{u} &= \int_{\Omega_e} {\M{B}^e}^T\bigg(
    2\mu_0  {\M{F^e}^T}\Big(\M{F^e}^T \M{F^e} -\M{C^*}\Big)    -   {\M{B}^e}^T \Vhatv \M{S}^*
    -
    \mu_0  {\M{B}^e}^T  \Vhatv {\M{B}^e}  \Vhatv {\M{F}^e}^T
    \bigg)     \td \Omega
    =0
    \\ 
    \M R_{\lambda} &= \int_{\Omega_e}
    \mu_0 {\M{B}^e}^T  {\M{F}^e}  {\M{B}^e}\Vhatv {\M{F}^e}^T
     +
    {\M{B}^e}^T  {\M{F}^e}  \M{S}^*
    \td \Omega
    - \M f^e = 0
\end{aligned}
\end{equation}
The solution of the system~\reff{eq:variSCsystemRfem} requires an iterative scheme, typically a Newton-Rhapson iteration. The  resulting finite element expressions are given in Appendix~2.

\section{Recording of the material data sets}\label{sec:rve}
The material data sets required for the DD-FEM are gained from systematic computations of the representative microscopic foam volumes. Hereby we expect the results to be physically meaningful, e.g., symmetric and regular. Also,
the specific choice of strain-stress tuples  $\V z_i$
is of minor importance; they can be converted into each other using the common relations of continuums mechanics. Therefore the data sets
\begin{align*}
  \mathcal{D}_F= \{(\T F,\T P)_{i}, i=1,\dots, n\}
  \qquad\text{and}\qquad
  \mathcal{D}_C= \{(\T C,\T S)_{i}, i=1,\dots, n\}
  \qquad
\end{align*}
or its small strain equivalent $\mathcal{D}_\epsilon= \{(\T\epsilon,\T \sigma)_{i}, i=1,\dots, n\}$
are treated equally and denoted as $\mathcal{D}$ subsequently.

\subsection{Generation of stochastic RVEs for open-cell foam}\label{subsec:rve}

\begin{figure}  
\includegraphics[width=\textwidth]{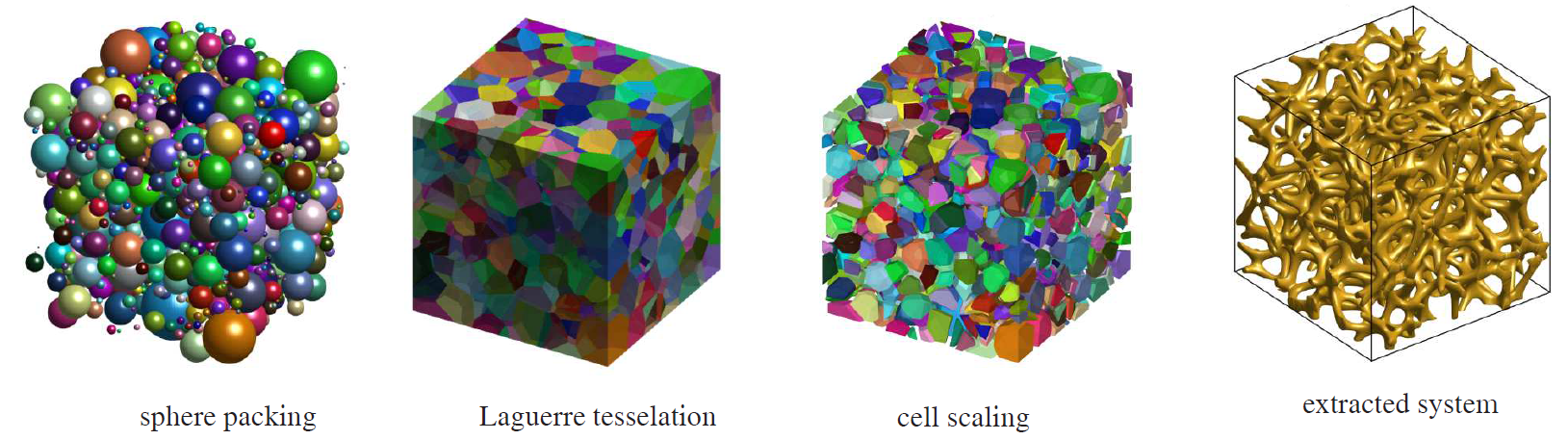}
\caption{Procedure to generate a representative stochastic foam volume: sphere radius and location are determined by stochastic variables. A Laguerre tesselation assigns a domain to every sphere. Then the cells are scaled to a given volume ratio and the skeleton is extracted afterwards.}\label{fig:rve:generation}
\end{figure}
Elastic open-cell foams are rubbery  materials with  a low relative density $R=\varrho_\text{foam}/\varrho_\text{elastomer} \leq \nicefrac{1}{3}$. Their mechanical behavior is determined by the matrix material and the cellular microstructure, consisting of a network of ligaments connected at junctions (vertices). An accurate description relies on  the real foam's geometry, whose topological characteristics are gained from computed tomography scans for example, cf.~\cite{Buchen_etalGAMM}. Assuming these characteristics, e.g. the pore volume fraction, the size distribution, the coefficient of variation, and the anisotropy factor, to be known, the following procedure is chosen to generate the corresponding RVEs.

At first, random sphere distributions with a collective rearrangement of the spheres are used to build a dense isotropic packing, see the left image of Fig.~\ref{fig:rve:generation}. A force-biased packing algorithm \cite{bargiel1991c} yields a very efficient procedure to attain the desired arrangement.  At next, a Laguerre tessellation, which is a  Voronoi tessellation weighted with the sphere radius \cite{liebscher2015laguerre}, is used to partition the volume into subregions, see the second image of Fig.~\ref{fig:rve:generation}. Then, the effective density of the foam is gained by a scaling of the cells, see the third image of Fig.~\ref{fig:rve:generation}. Finally, the edges need to be extracted.  Originally they had been shared by three cells and through the scaling they have now a hexagonal cross section (a triangular shape with cut corners). These polygonal struts are  the foam's ligaments and still need to get  realistic  cross sections. The adjacent struts are joined in the junctions and because this results in kinks and internal corners a smoothing algorithm is applied. We make use of a spline-based algorithm proposed in \cite{loop1994smooth}. The result can be seen in the right image of Fig.~\ref{fig:rve:generation}.

\begin{figure}[htb!]
\centering
\includegraphics[width=0.45\textwidth]{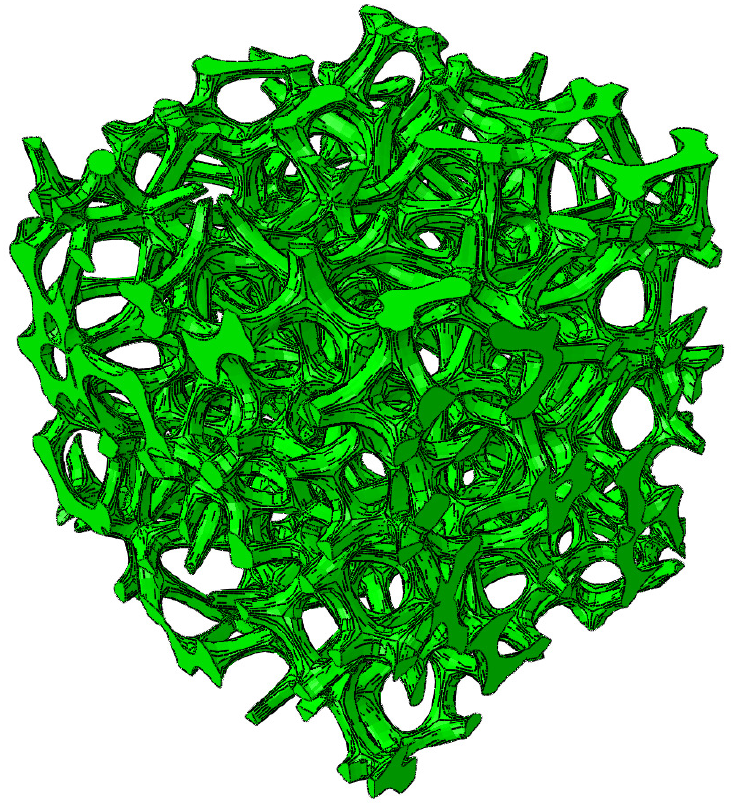}
\includegraphics[width=0.45\textwidth]{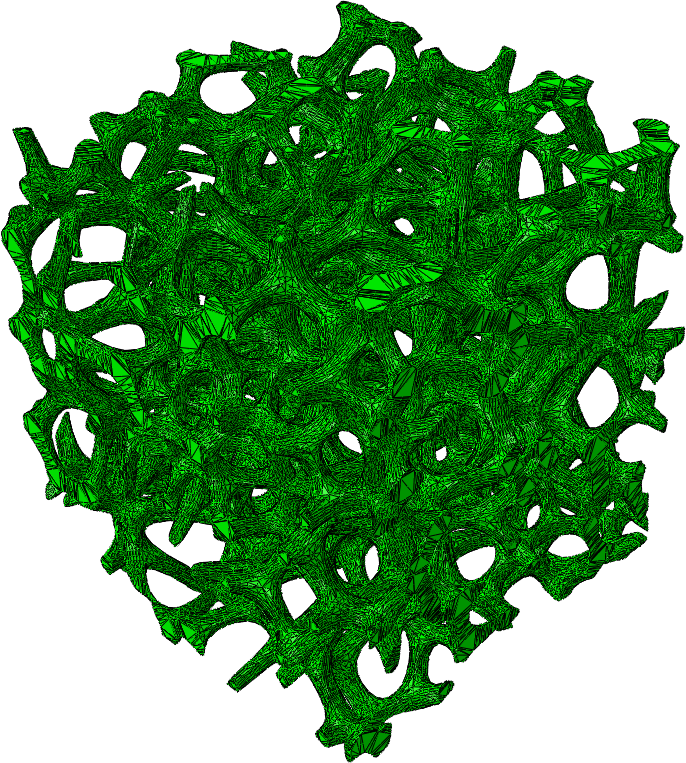}
\caption{RVE of a foam with 100 log-normally distributed  pores, a coefficient of variation of $0.5$, and a relative density of $0.15$ (left); its finite element model consists of $708\,372$  tetrahedral elements (right).}\label{fig:rve_bsp}
\end{figure}

For a finite element analysis the resulting RVE still needs to be meshed.
We employ here linear tetragonal solid elements; an example RVE of size $1000\times 1000\times 1000\,$\textmu m
can be seen in Fig.~\ref{fig:rve_bsp}.
The displayed RVE is meshed with 708\,372 elements and a single linear computation with periodic  boundary condition takes around 10 minutes. This effort is important because many -- linear and non-linear -- computations need to be conducted to obtain the required data sets.

\subsection{Homogenized microscopic data set}
To generate a data set that describes the homogenized material behavior of the foam, a deformation $\bar{\tens{F}}$ of the RVE is prescribed and the stress field is computed with a linear or
non-linear FEM. The homogenized stresses $\bar{\T P}$ or $\bar{\T S}$ are derived by
\begin{align*}
\bar{\T S}  &=\frac{1}{|V|}\int_{V} \T S(\V X) \td V\,,
\end{align*}
i.e. for all integration points the stress components multiplied with the corresponding volume are summarized and divided by the RVE's volume $V$.
The data sets, which are used as an input for the DD-FEM, are  then the collections of all $(\bar{\tens{F}},\bar{\tens{P}})$ tuples or, with $\bar{\tens{C}}=\bar{\tens{F}}^T\bar{\tens{F}}$ all  $(\bar{\tens{C}},\bar{\tens{S}})$ tuples.
Naturally, the question arises of how many computations need to be conducted to obtain a data set that sufficiently describes a foam's material behavior. Our experience suggests a minimum number of $n=50-100$ data tuples per stress component, which results in the computation of $n^6$ tuples for three dimensions at least. In the following, we discuss how we gain these data collections  efficiently in cases of different material characteristics.

\paragraph{\textbf{Case A: }non-linear and anisotropic material}\quad\\
For the general situation of a direction dependent and non-linear material response, e.g. a viscoelastic foam with elongated pores in one direction, simplifications are difficult. Here only a brute force approach of one RVE computation for every $(\bar{\tens{F}},\bar{\tens{P}})$ or $(\bar{\tens{C}},\bar{\tens{S}})$ tuple  gives the wanted data set $\mathcal{D}$. Such a strategy is hardly feasible with the RVEs described above, only a simpler beam discretization   may help.  Additionally, it needs to be considered whether these volume elements are still representative.

\begin{figure} 
\centering
\includegraphics[width=0.75\textwidth]{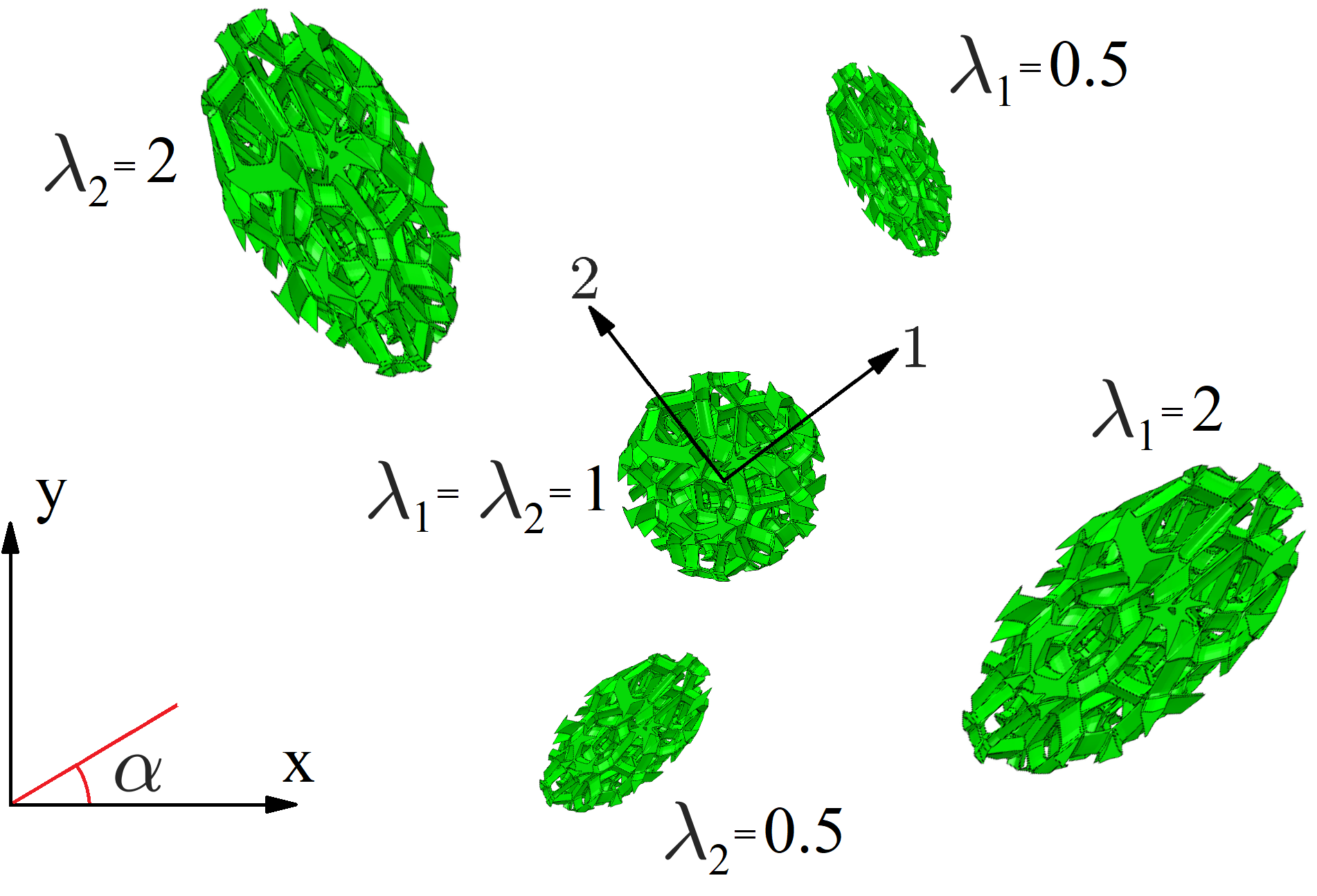}
\caption{Elongation in principle stretches and rotation to the reference coordinate system for d=2}
\label{fig:Eigen}
\end{figure}

\paragraph{\textbf{Case B:} non-linear and isotropic material}\quad\\
For an isotropic material, i.e. when the microscopic response is non-linear but does not depend on the loading direction, we gain the possibility to rotate the RVE's coordinate system, see Fig.~\ref{fig:Eigen}. Our strategy is here to sample discrete values of the principal stretches $\lambda_\alpha$, $\alpha=1 \dots $d, and then transform the resulting principal stresses,
\begin{align*}
  \bar{\tens{S}} = \tens{Q}  \bar{\tens{S}}^{\lambda_\alpha} \tens{Q}^T
\end{align*}
where $\T Q \in \SO3$ is the corresponding three-dimensional rotation tensor composed of the rotations around the coordinate axes, \cite{KirchdoerferOrtiz2016}.
With a stepwise rotation around these axes all deformation states can be mapped and summarized in $\mathcal{D}$. For $n^6$ data then $n^3$ computations are needed. 

\begin{figure}[htb!]
\includegraphics[width=1\textwidth]{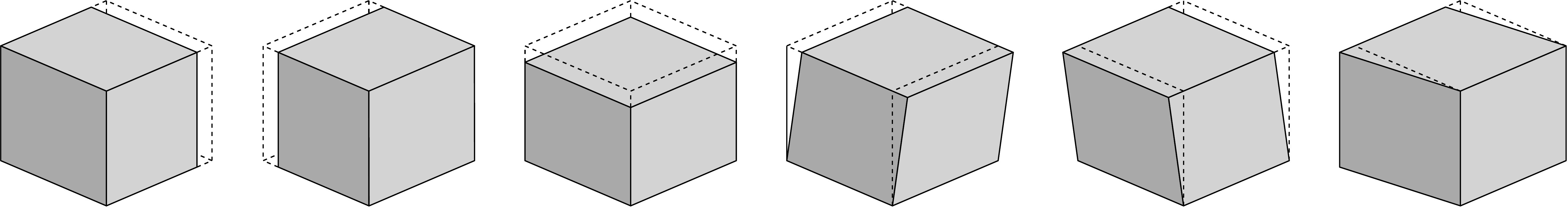}
\caption{Unit load deformations of an RVE for direction dependent linear material. }\label{rve_zustande}
\end{figure}

\paragraph{\textbf{Case C:} linear and anisotropic material}\quad\\
The computation eases significantly when  the applied deformation is small and the microscopic response is linear. Without isotropy of the material still RVE computations with loads in every distinct direction need to be performed but the results now can be superposed 
\begin{align*}
  (\bar{\tens{C}},\bar{\tens{S}})_{k} = a (\bar{\tens{C}},\bar{\tens{S}})_{i} + b (\bar{\tens{C}},\bar{\tens{S}})_{j}
\end{align*}
with $a,b \in \IR$. Conveniently, unit loads as displayed in Fig.~\ref{rve_zustande} are computed and evaluated. This strategy leads to a enormous decrease of computational effort and will later be employed for the small deformation data, see \ref{appendix3}.

\paragraph{\textbf{Case D:} linear and isotropic material}\quad\\
In the simplest case of isotropic, linear (elastic) material the loading scenarios simplify even more. 
The material's isotropy  makes it possible to describe the material by only two states, namely one of the  three elongations and one of the three shear states from Fig.~\ref{rve_zustande}.
All other data tuples can be gained  by  linear combinations and  rotations of the coordinate system.

Here we remark that for these simplifications 
we assume the RVEs to give the same averaged response, i.e., we neglect the noise induced by the stochastic generation. The uncertainty in the data will be subject of a subsequent work.

\subsection{Multi-level method}
The computational effort of a data-driven multiscale analysis is enormous. As outlined above, for a non-linear microscopic material response many RVEs need to be  computed to gain the required data. Additionally, data-driven computations are numerically costly due to the search for the optimal data points and, furthermore, for a macroscopic non-linear kinematic the solution requires an iterative solution procedure. To reduce these computational costs, we apply here a multi-level method which was introduced in \cite{korzeniowski2021multi}.
In this procedure we compute the solution on smaller subsets of the original set 
$\mathcal{D}_l \subset \mathcal{D}$. After a computation at level $l$ with data set $\mathcal{D}_l$, relevant  data from the total set $\mathcal{D}$ are added  to the next set $\mathcal{D}_{l+1}$. Here we go a step further than in \cite{korzeniowski2021multi}, and use the multi-level method to identify regions where more RVE calculations are needed and then calculate additional deformation states on the fly.

\begin{figure}
\centering
\includegraphics[width=0.6\textwidth]{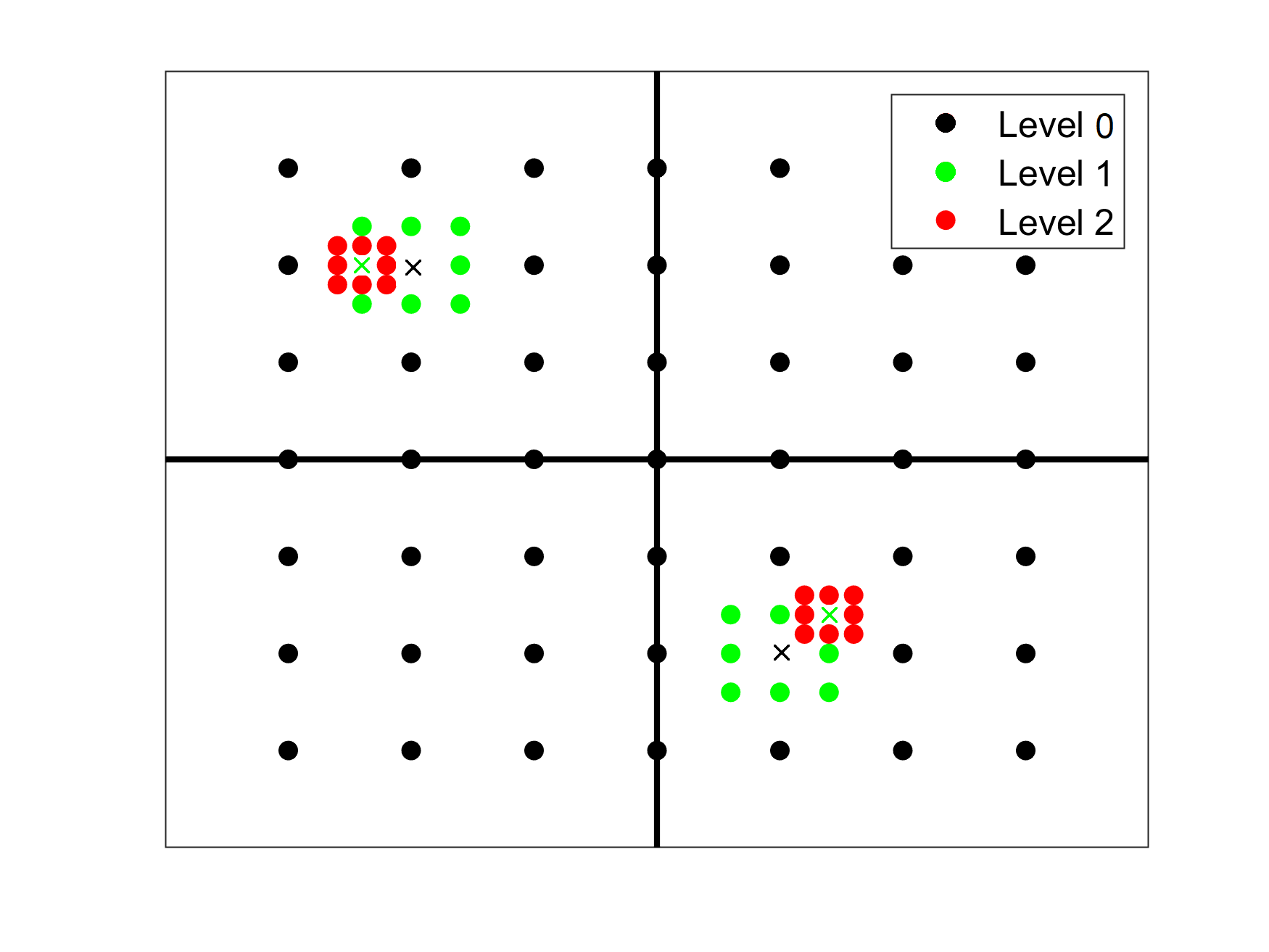}
\caption{Illustration of the multi-level method for a simple state space: The black data points symbolize the input set of the initial level $\mathcal{D}_0 $; the crosses are the data points actually assigned to the material points. The green and red points mark the data of the first and second refinement level, $\mathcal{D}_1$ and $\mathcal{D}_2$, respectively. }\label{fig:MLrefinement}
\end{figure}

\begin{figure}
\centering
\includegraphics[width=0.6\textwidth]{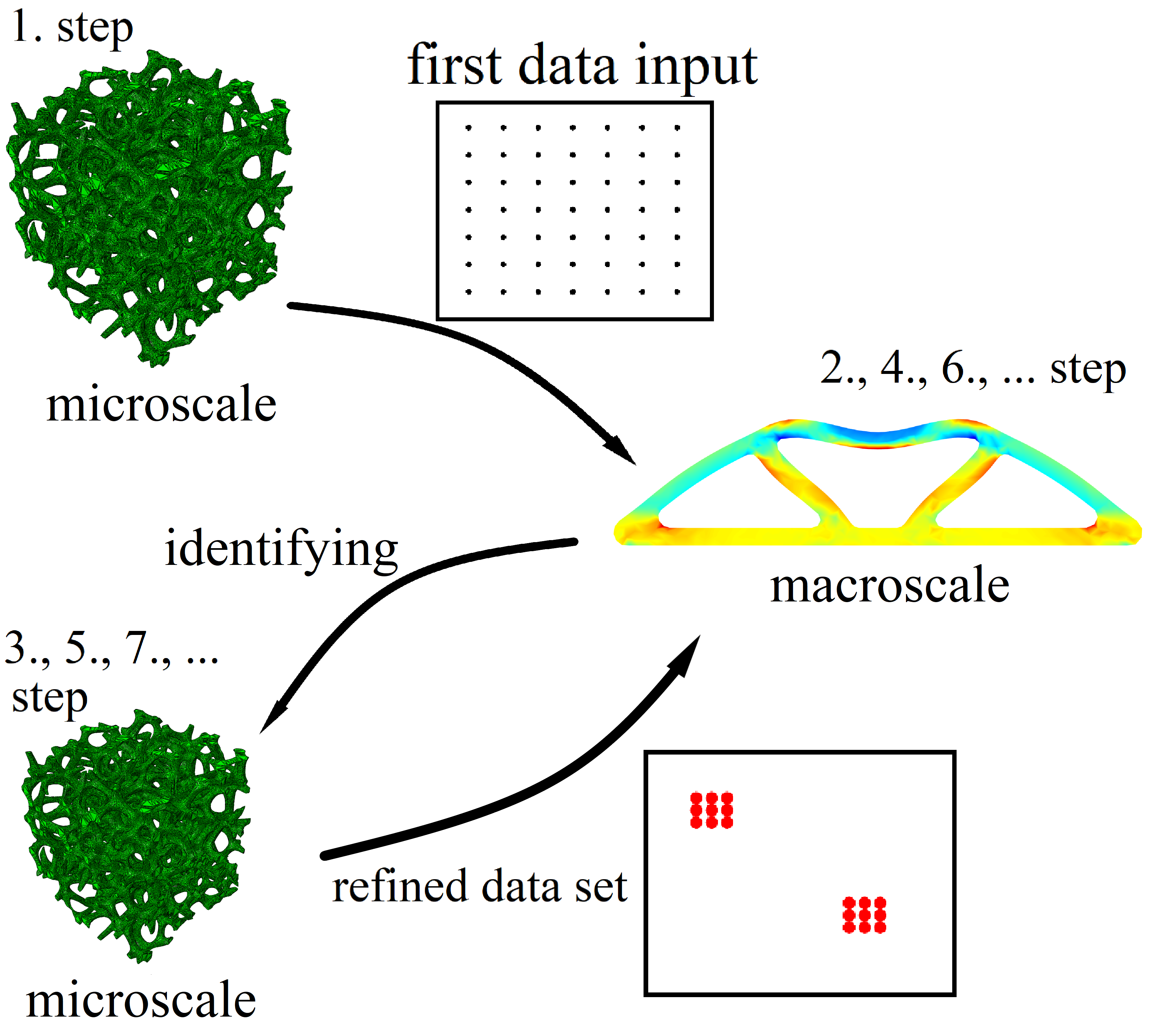}
\caption{Illustration of the multi-level approach. A coarse data set is generated by RVE simulations first. A macroscopic data-driven simulation afterwards identifies region where more precise data is needed. Additional RVE simulations are conducted to generate new data in the desired regions. The latter two steps can be repeated as often as desired. }\label{fig:ml_fluss}
\end{figure}

Fig.~\ref{fig:MLrefinement} shows exemplary three  data sets on a (simplified) state space. The aim is to cover the space of possible states (black square)  as good as possible by data tuples (dots). Instead of a high density of data, at first only a coarse set $\mathcal{D}_0 $ is created  through  RVE calculations (black dots, level 0). This set $\mathcal{D}_0$ is used for a first DD-FEM of the structure. In this  computation, the approximate data of relevance are identified by the data tuples which are assigned to the material points (black $\times$).  Then  the data set is adaptively refined and the data-driven FEM is repeated with a set $\mathcal{D}_1 $. In Fig.~\ref{fig:MLrefinement} this is exemplarily illustrated by the two green $\times$, marking the selected data of set $\mathcal{D}_1$. Around those two assigned data tuples  the data grid is refined.  This leads to the data set $\mathcal{D}_2$ (red dots).  This procedure can be repeated as often as needed to achieve a sufficient accuracy. The adaptive computational strategy for the multi-scale simulation is illustrated in Fig.~\ref{fig:ml_fluss}.

\section{Numerical Examples}\label{sec:numeric}
To demonstrate the different DD-FEM approaches, we start our computations  with a  parametric study of a simple rod. Then
we show the applicability of the method for the computation of engineering components by means of a polyurethane rubber sealing.

\subsection{Example 1: Rod under tension }\label{sec:stab}
At first we investigate the influence of different data sets together with a linear or a non-linear kinematic.
For these parametric studies we consider a one-dimensional rod of length $l=100\,$mm and cross-section $A=1\,$mm$^2$ under uniaxial tension. Assuming incompressibility of the material, the  state of deformation is given by
\begin{align*}
\T F=\begin{pmatrix}
\lambda_1 & 0 & 0 \\ 0 & \lambda_1^{-1/2} & 0\\ 0 &0 &\lambda_1^{-1/2}
\end{pmatrix},
\end{align*}
which means we can uniquely describe it with the lateral stretch $\lambda_1$. In this example,  we do not obtain the material data sets from any particular foam RVE because this would not be incompressible. Instead we generate the data artificially. These data sets are then used in the two algorithms of Section~\ref{sec:basics}.


\paragraph{Artificial material data generation}
We consider the lateral stretch $\lambda_1$ to be given and assign the corresponding first Piola-Kirchhoff stress $P_1$. Three sets of data are generated, assuming one proportional stretch-stress relation and two different strain energy densities,
whereby in the latter cases we assume the  continuum mechanics relation $\T S = \tens{F}^{-1}\tens{P} $ to be valid.

\begin{figure}
\centering
\includegraphics[width=0.65\textwidth]{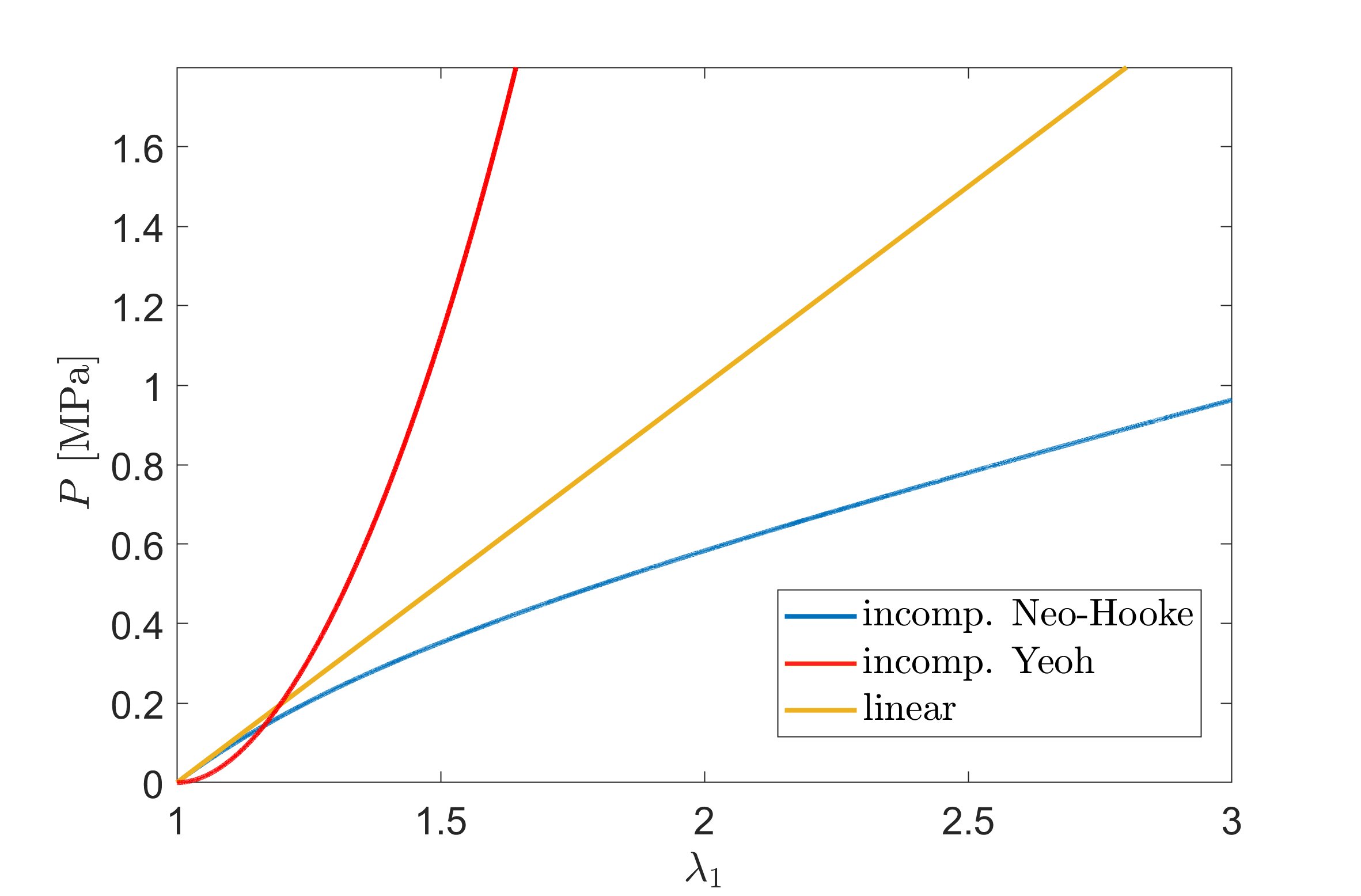}
\caption{Data of the three sets for the one-dimensional study.}\label{fig:1d_data}
\end{figure}

\begin{itemize}
\item[(i)] Linear data: Assuming proportionality between $\lambda_1$ and $P_1$ as well as $\lambda_1^2$ and $S_1$ we gain the following data sets:
\begin{align*}
( F,  P)_{i}&=\left(\lambda_1,\, c_1\lambda_1\right)_{i} 
\\
( C,  S)_{i}&=\left(\lambda_1^2,\, c_1\lambda_1^2\right)_{i} \qquad \text{for }i=1,\dots,n
\end{align*}
with material constant $c_1= 1\,$MPa.
%
\item[(ii)] Neo-Hookean-like data: Let the data derive from an incompressible material's strain energy density of the form
\begin{align*}
W&=c_1(\tr (\T F^T \T F)-3)+p(\det \T F -1)
\end{align*}
where $p$ is a Lagrange multiplier enforcing incompressibility, we deduce the two data sets from the corresponding analytical solution, cf. \cite{reppel2013elastic}.
\begin{align*}
( F,  P)_{i}&=\left(\lambda_1,\, 2c_1(\lambda_1-\lambda_1^{-2})\right)_{i} 
\\
( C,  S)_{i}&=\left(\lambda_1^2,\, 2c_1(1-\lambda_1^{-3})\right)_{i} \qquad \text{for }i=1,\dots,n
\end{align*}
The single material constant is here $c_1 = \nicefrac16\,$MPa.
\item[(iii)] Yeoh-like data: Using a strongly non-linear strain energy density of the form
\begin{align*}
W&=c_1(\tr (\T F^T\T F)-3) + c_3(\tr (\T F^T\T F)-3)^3 +p(\det \T F -1)
\end{align*}
we derive in the same way the data tuples, cf. \cite{yeoh1990characterization}
\begin{align*}
( F,  P)_{i}&=\left(\lambda_1,\, 2(\lambda_1-\lambda_1^{-2})\left(c_1+3c_3(\lambda_1^{2}+2\lambda_1^{-1}-3)^2 \right)\right)_{i} 
\\
( C,  S)_{i}&=\left(\lambda_1^2,\, 2(1-\lambda_1^{-3})\left(c_1+3c_3(\lambda_1^{2}+2\lambda_1^{-1}-3)^2 \right)\right)_{i} \quad \text{for }i=1,\dots,n
\end{align*}
with constants $c_1 = \nicefrac16\,$MPa and $c_3=\nicefrac{1}{1000}\,$MPa.
\end{itemize}
The data of the three sets are plotted in Fig.~\ref{fig:1d_data}. Because we chose very  dense data sets with $n=10\,000$, single data points are not recognizable in the plot but the non-linearity can clearly be seen.

\begin{figure}
\centering
\includegraphics[width=0.55\textwidth]{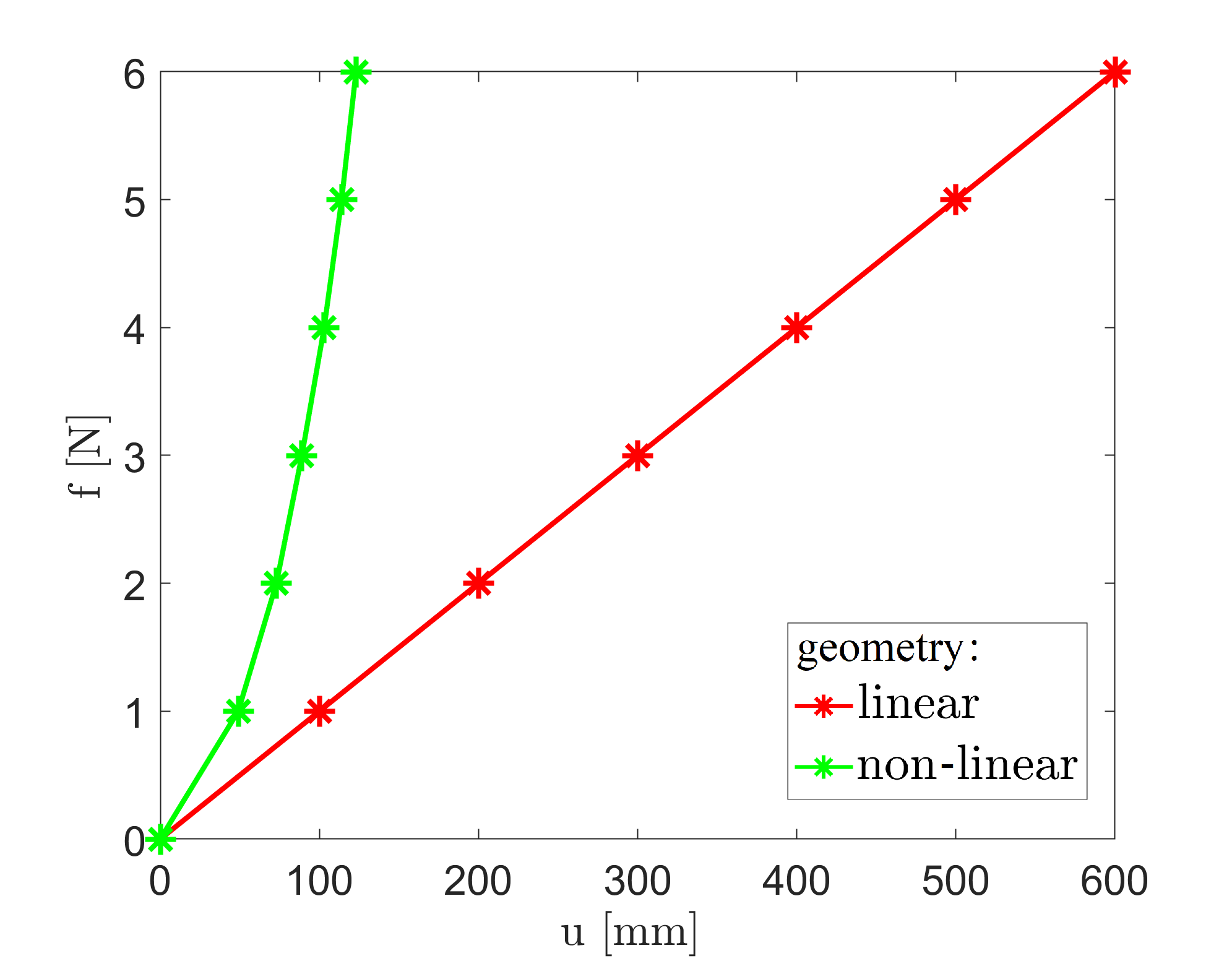}
\caption{Load-displacement curves for proportional material sets (i) computed with the linear $(\T P,\T F)$ formulation and the non-linear $(\T C, \T S)$ formulation.}\label{fig:stab_matlin}
\end{figure}

\begin{figure}
\centering
\includegraphics[width=0.48\textwidth]{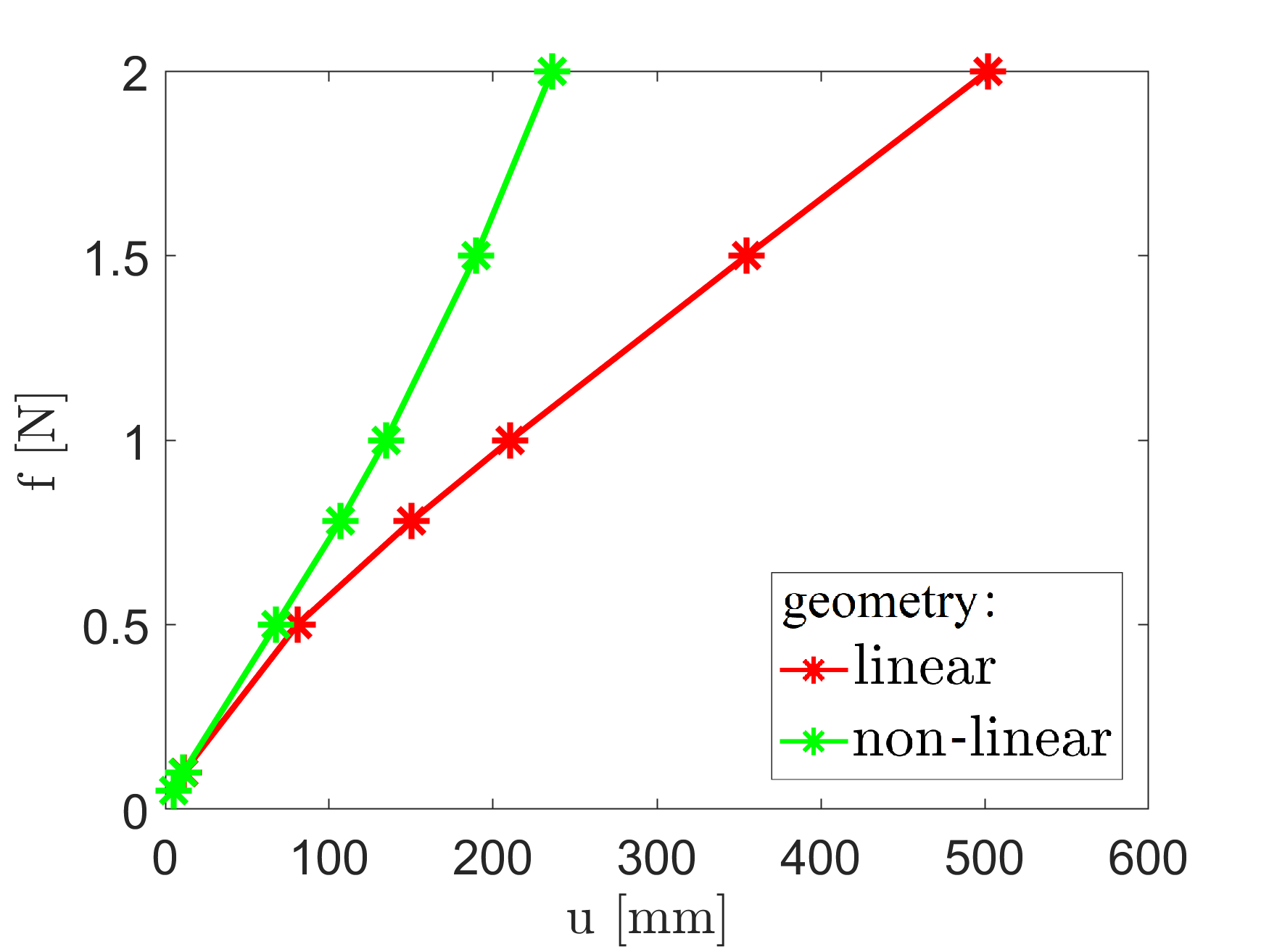}
\includegraphics[width=0.48\textwidth]{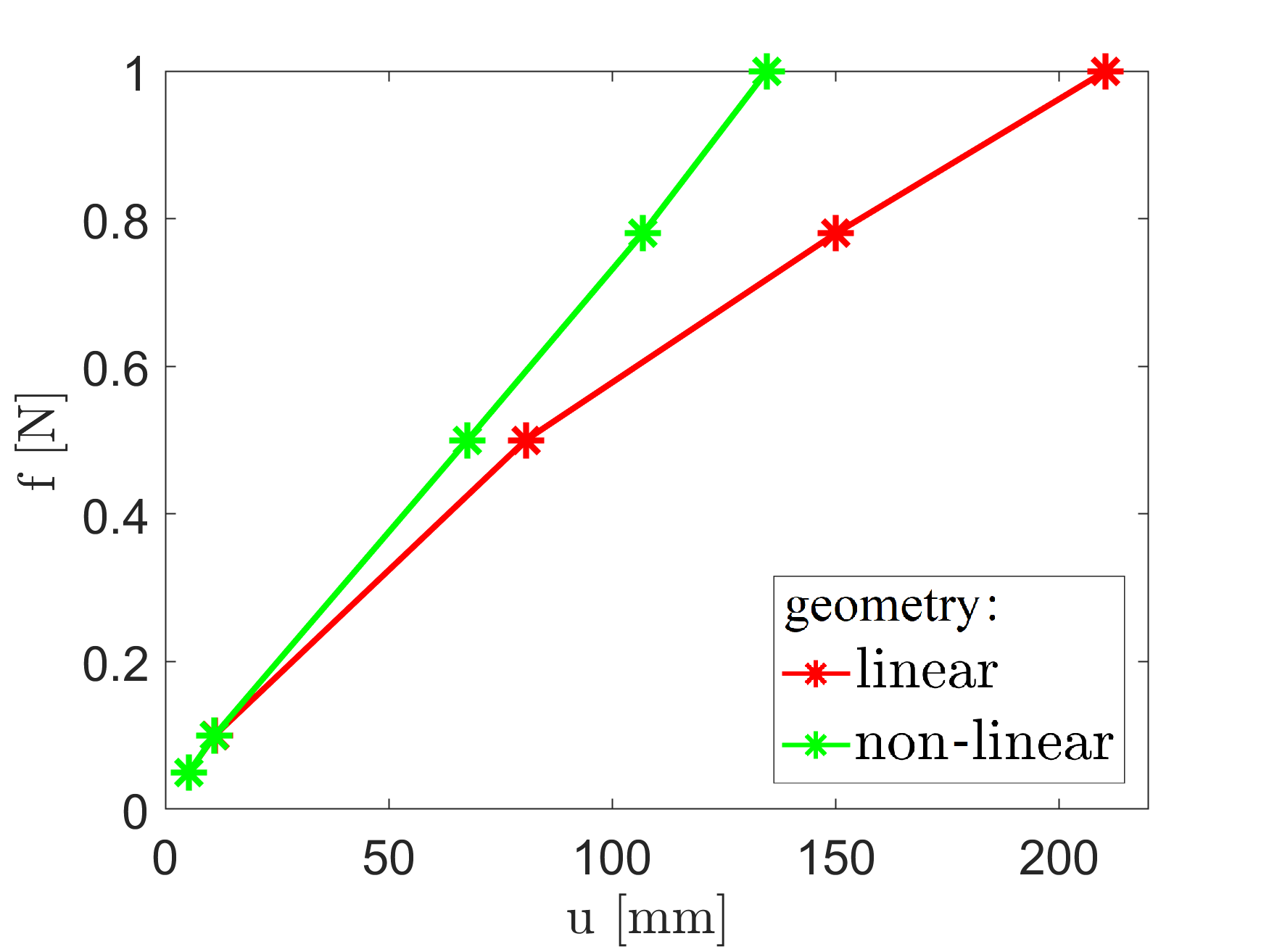}
\caption{Load-displacement curves for the non-linear material set (ii) computed with the linear $(\T P,\T F)$ formulation and the non-linear $(\T C, \T S)$ formulation; two different regimes with $500\%$ straining (left) and  $100\%$ straining (right) are shown.}\label{fig:matNlin_u}
\end{figure}

\begin{figure}
\centering
\includegraphics[width=0.48\textwidth]{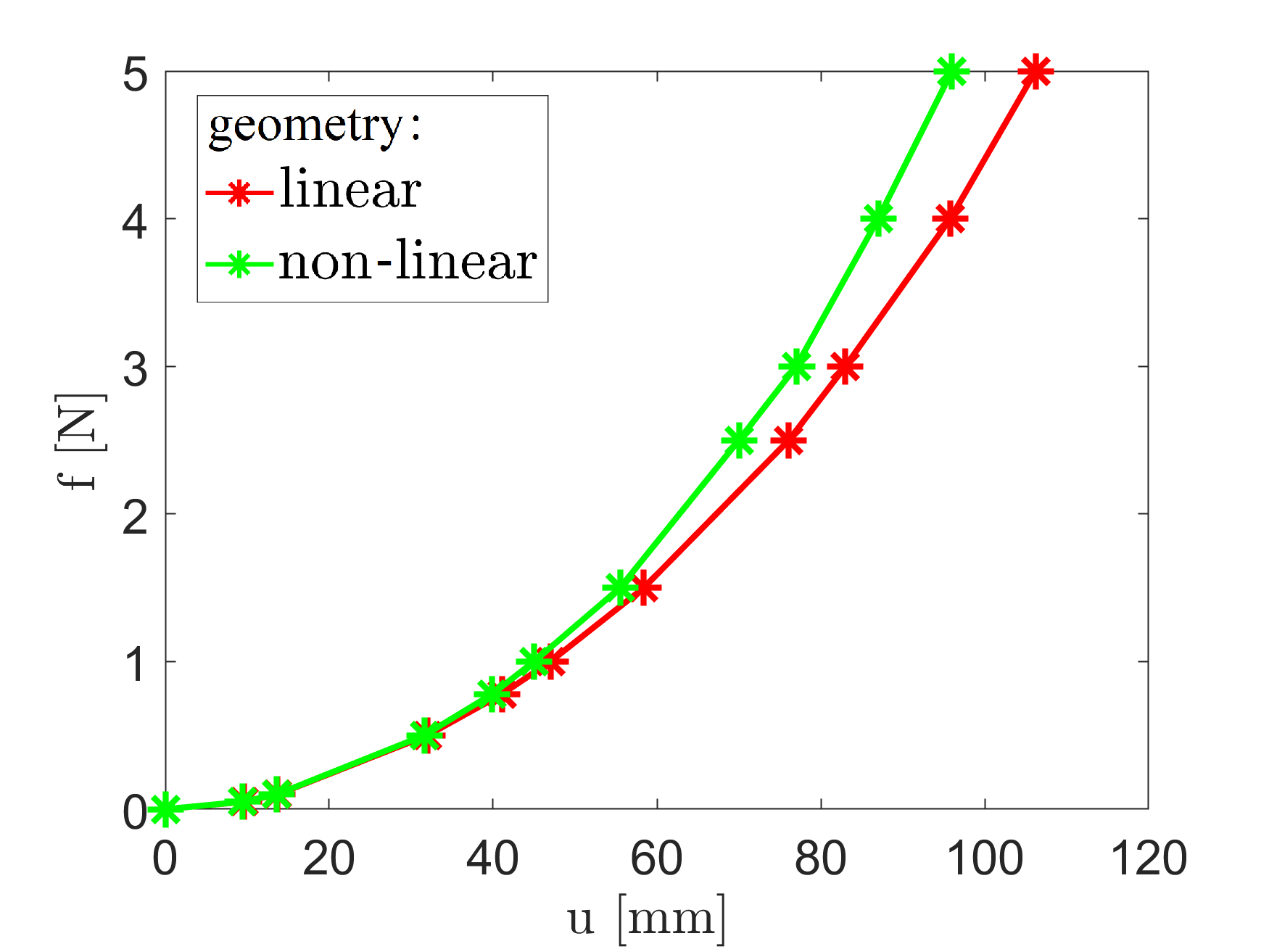}
\includegraphics[width=0.48\textwidth]{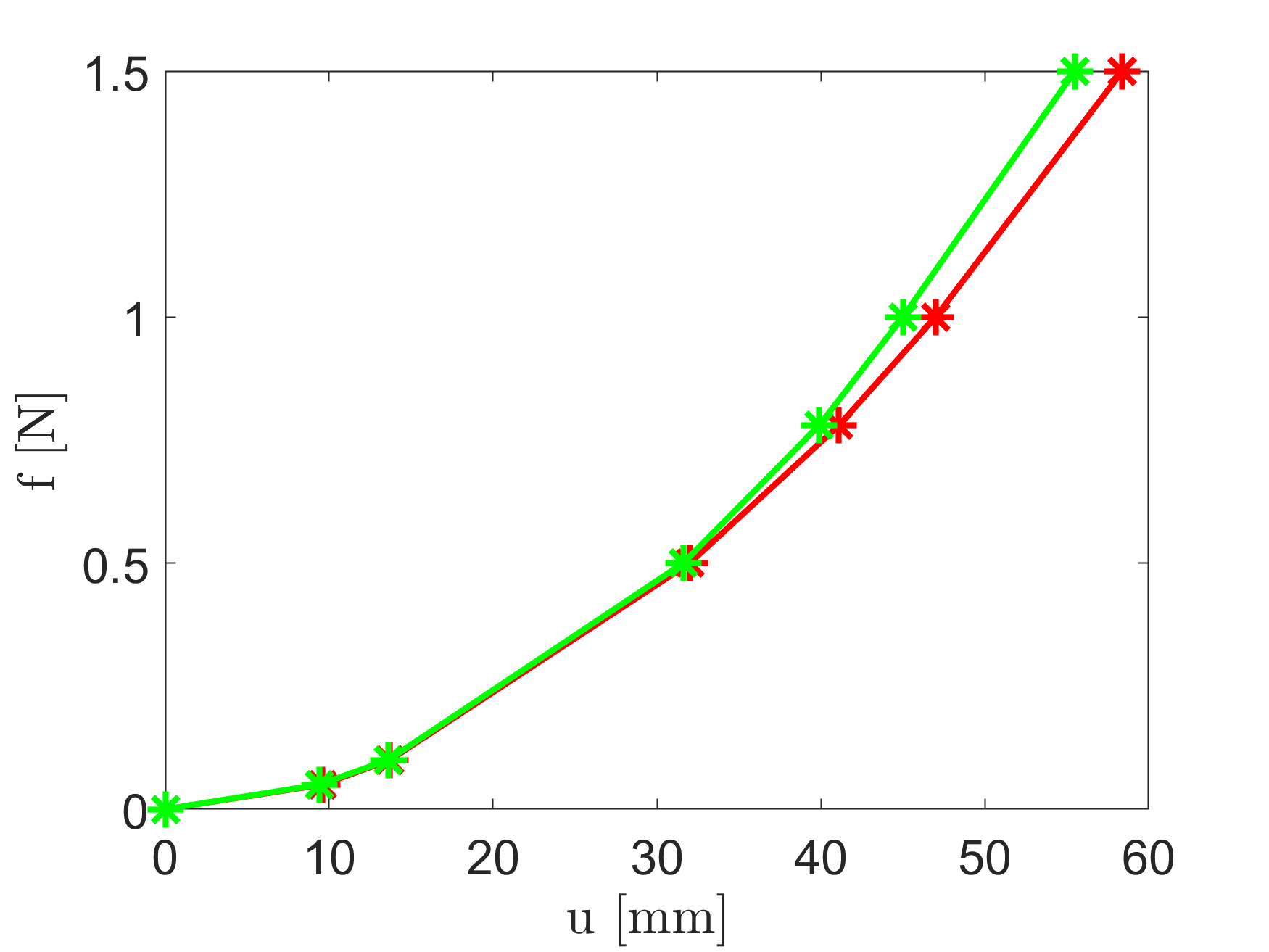}
\caption{Load-displacement curves for the strongly non-linear material set (iii) computed with the linear $(\T P,\T F)$ formulation and the non-linear $(\T C, \T S)$ formulation; two different regimes with $100\%$ straining (left) and  $60\%$ straining (right) are shown.}
\label{fig:u_YH_incompre}
\end{figure}

\paragraph{Comparative study}
We are interested in the force-displacement relation for a homogenously elongated bar and evaluate the maximum displacement at the bars end u$=(\lambda_1-1) l$.  In 
Fig.~\ref{fig:stab_matlin} the corresponding force-displacement curves are plotted for the proportional material data of set (i). Clearly, computing the DD-FEM with the linear kinematic of Section~\ref{sec:FB} we obtain a linear u-f-relation, whereas with the non-linear kinematic of Section~\ref{sec:DD_CS} the
maximum displacement is reduced. The same situation is displayed in Fig.~\ref{fig:matNlin_u} for the DD-FEM computation with the sub-linear Neo-Hookean set (ii)
and in Fig.~\ref{fig:u_YH_incompre} for the strongly non-linear data set (iii).

In case (ii) two effects work against each other: The non-linear data are softer for larger displacements but the non-linear kinematics stiffens the model. In other words, while the kinematic gives a higher stretch at the same displacement, the proportionality of model (ii)  gives a lower stress at the same stretch. The difference between the two approaches is about $14\%$, $21\%$ and $41\%$ at $\lambda_1=1.5$, $\lambda_1=2$ and $\lambda_1=3$, respectively.

In  Fig.~\ref{fig:u_YH_incompre} the strong non-linearity of the material dominates the situation. The  u-f-curves are similar
but the two effects almost balance.
The linear or non-linear kinematic hardly makes a difference here because the displacements are also significantly smaller than before.
The relative difference between both DD-FEM computations is $8\%$ and $19\%$ at  $\lambda_1=1.5$ and $\lambda_1=2$, respectively.

Summarizing we state that in the DD-FEM the structure of the material data set dominates the solution. For moderate straining and considering some expected noise in the available data sets, a non-linear material behavior can be mapped with a linear kinematic sufficiently accurate. 
This result is important because a non-linear kinematic with a Newton-Raphson iteration and a data search in every step as described in \ref{appendixCS} is by several factors more expensive than a linear DD-FEM computation.

\clearpage
\subsection{Example 2: A rubber sealing in a plane strain state}\label{sec:exa2}
\begin{figure}[htb!]
\centering
\includegraphics[width=0.35\textwidth]{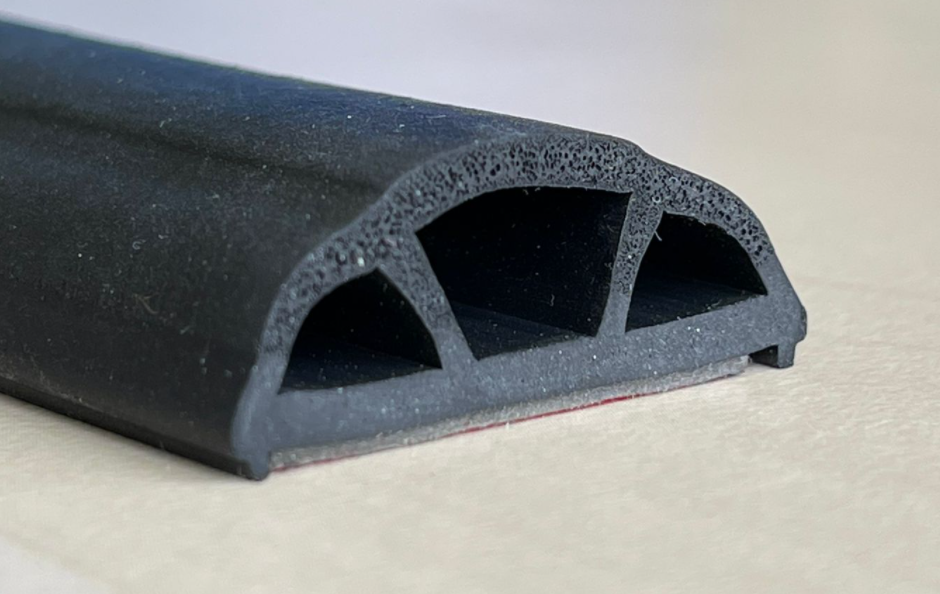}\hfill
\includegraphics[width=0.6\textwidth]{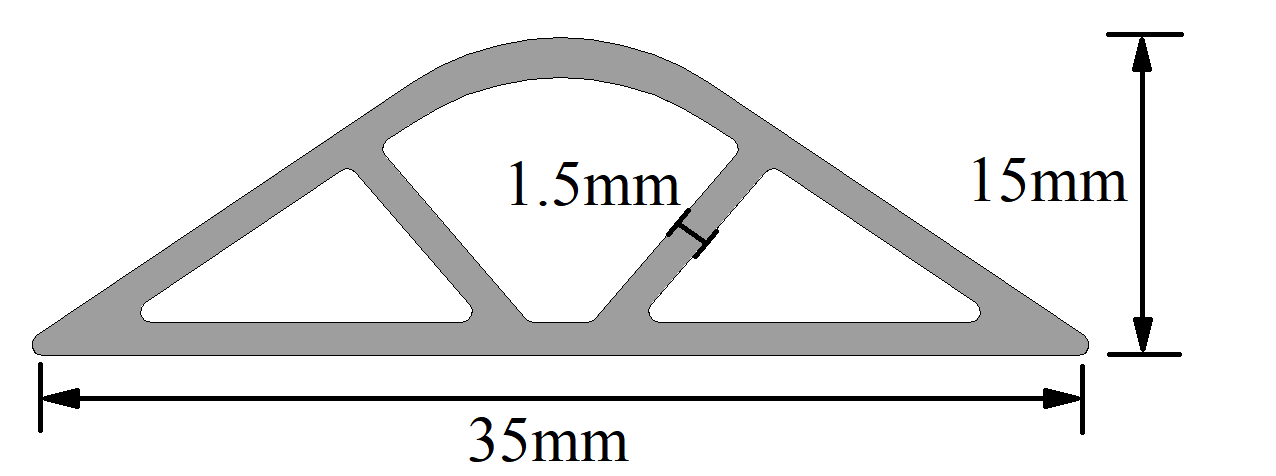}
\caption{Geometry of the rubber sealing  }\label{fig:_dichtung}
\end{figure}

Now we consider a  typical engineering application, namely a car door sealing made of a foamy rubber material. The computed geometry is displayed in Fig.~\ref{fig:_dichtung}.

We start with a somewhat artificial situation of a plane approximation and  two-dimensional foam data.
This corresponding  RVE is a simple $7 \times 7\,$mm$^2$ square with 4 regularly arranged pores of radius $2\,$mm  and meshed with 2601 finite elements.
The RVE matrix  is a Neo-Hookean material extended to the compressible range, $W=c_1 (\tr(\T C) -3 )+\nicefrac{1}{D_1}(J-1)^2$,
with parameters of polyurethane, 
$c_{1}=3.85\,$MPa and $D_1=0.12$ MPa$^{-1}$. The computation is performed with the commercial program \textsc{Abaqus}. 

\begin{figure}[htb!]
\centering
\includegraphics[width=0.4\textwidth]{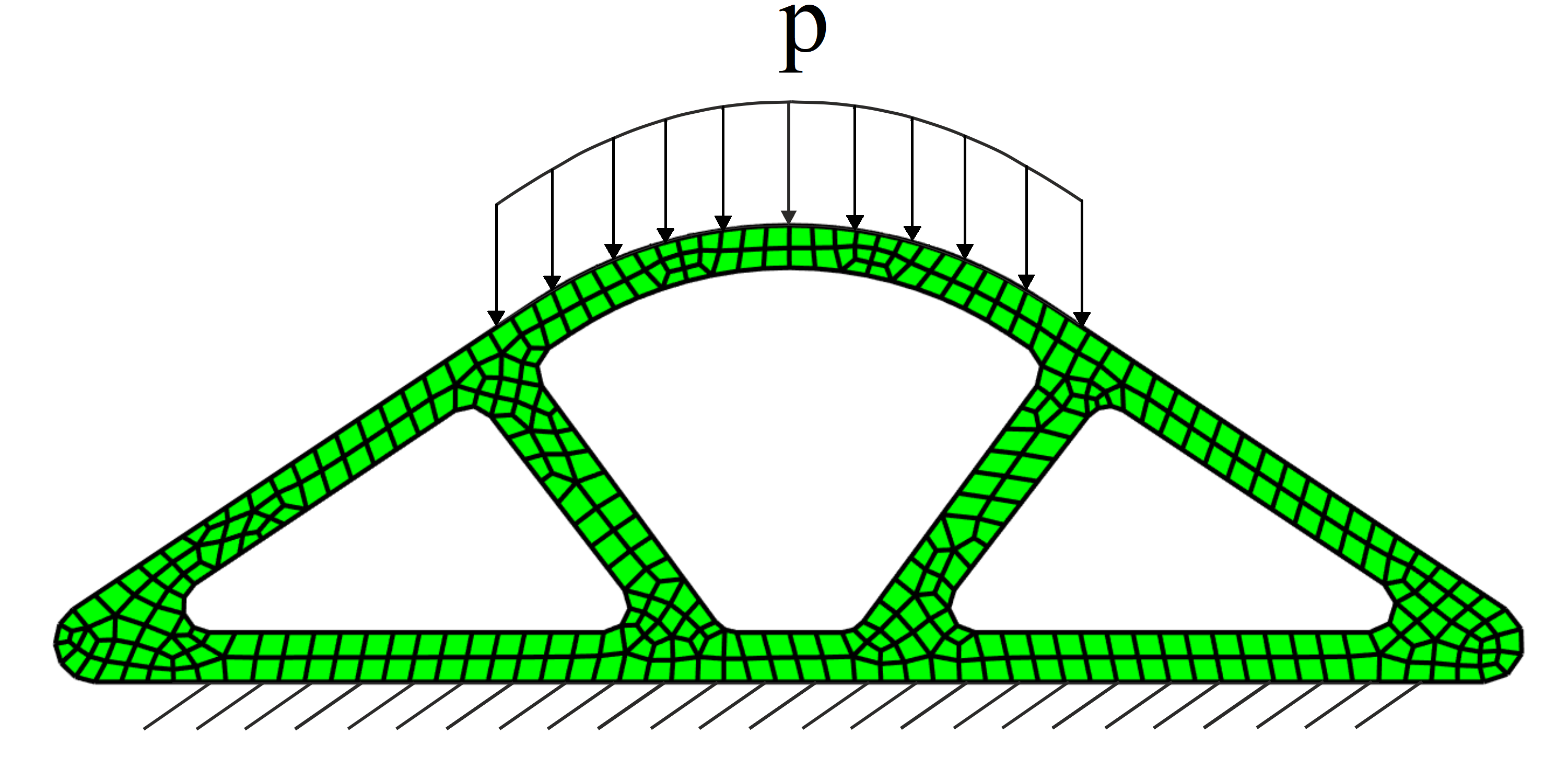}
\includegraphics[width=0.58\textwidth]{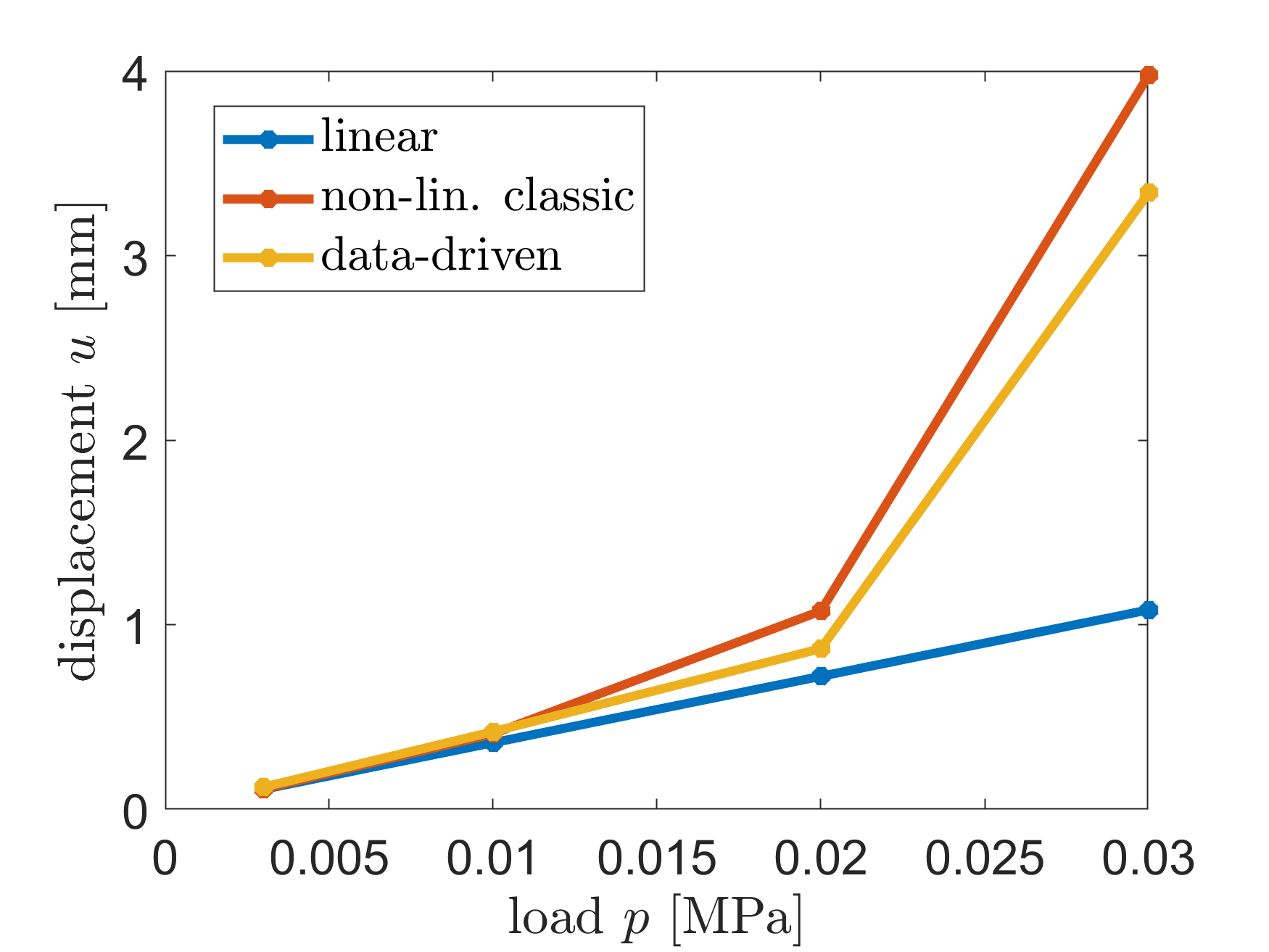}
\caption{Mesh and boundary conditions of the rubber sealing on the left side. On the right side the load-displacement curve is displayed for the three simulations.}\label{fig:examp2_bc}
\end{figure}

\paragraph{Macroscopic finite element model}
The mesh of the sealing, its loading and boundary conditions are shown on left-hand side of Fig.~\ref{fig:examp2_bc}.
The component is fixed in the $z$-direction. For the distributed surface load $p$ four different values are used, namely $p=3\,$kPa, $p=10\,$kPa, $p=20\,$kPa and $p=30\,$kPa.
For comparison, linear (material and kinematic) and non-linear (material and kinematic) finite element analyses based on the classical material models were also performed.

\paragraph{Generation of the material data set}
For data generation the principal stretches $\lambda_1$ and $\lambda_2$ were varied in 51 steps between $0.8 \dots  1.2$ which lead to 2601 non-linear finite element computations of the RVE. The resulting data set has $2601\cdot 73=189\,873$ data tuples  which provide  in our data-driven simulation  the initial set $\mathcal{D}_0$ of the multi-level approach.

\paragraph{Data-driven finite element analysis}
The data-driven computations were run on three levels of refined data sets.
In Table \ref{table:MLex2} the size of the input data set and the number of data tuples which are actually assigned to a material point (solution set $\mathcal{S}$)
are displayed for the simulation with the highest load $p=30\,$kPa.

On the right-hand side of Fig.~\ref{fig:examp2_bc} the load-displacement curves are shown for the data-driven solution and the two classic computations. Recorded is the
displacement of the center top nodes where we have the largest displacement.
%
The data-driven solution, computed with the non-linear material data but the linearized $(\T F,\T P)$ formulation of Section~\ref{sec:FB}, gives a somewhat stiffer
response than the fully non-linear Neo-Hookean model. The difference in the maximal displacement is about $15\%$ for moderate straining. 
As outlined in the previous example, a fully non-linear kinematic of the DD-FEM comes at the price of a much higher computational effort and
so we consider the linearized data-driven response here to describe  the deformation of the structure sufficiently.

\begin{figure}[htb!]
\centering
\includegraphics[width=0.88\textwidth]{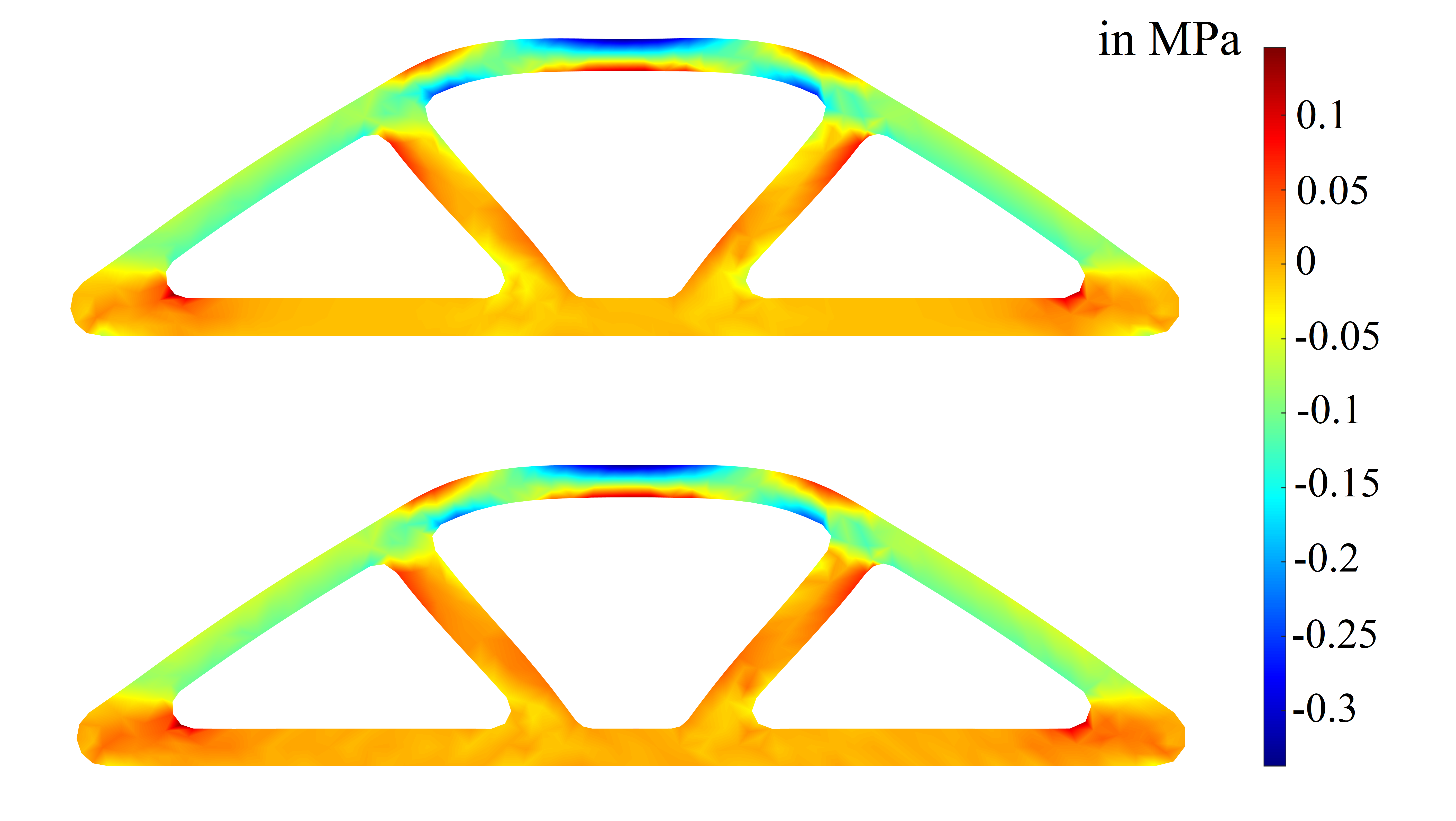}
\caption{Horizontal Cauchy stress $\sigma_{x}$ for a classic linear FEM solution (top) and the DD-FEM (bottom) for a surface load of $p=20\,$kPa. }\label{fig:examp2_stress}
\end{figure}

In Fig.~\ref{fig:examp2_stress} the horizontal Cauchy stress component $\sigma_x$ is displayed for $p=20\,$kPa and a good agreement of the classic FEM solution and the data-driven solution (level 2) can be seen.

\begin{table}
\centering
\ra{1.2}
\begin{tabular}{crc}\toprule
Level $l$  & data tuples in $\mathcal{D}_l$& data tuples in $\mathcal{S}_l$   \\ \hline
0 & 189.873 & 2.972    \\
1 & 80.244 & 3.197  \\
2 & 86.319 & 5.201  \\\hline
\bottomrule
\end{tabular}
\caption{Sizes of the input data and the number of data tuples describing the material of the multi level computation.  }
\label{table:MLex2}
\end{table}

\subsection{Example 3: Three-dimensional computation of the rubber sealing} \label{sec:exa3}
We proceed with the rubber sealing component of Section~\ref{sec:exa2} but compute now a fully three-dimensional deformation state with a different loading scenario. The  material is  described with  microscopic RVEs  of polyurethane foam.

\begin{figure}[htb!]
\centering
\includegraphics[width=0.48\textwidth]{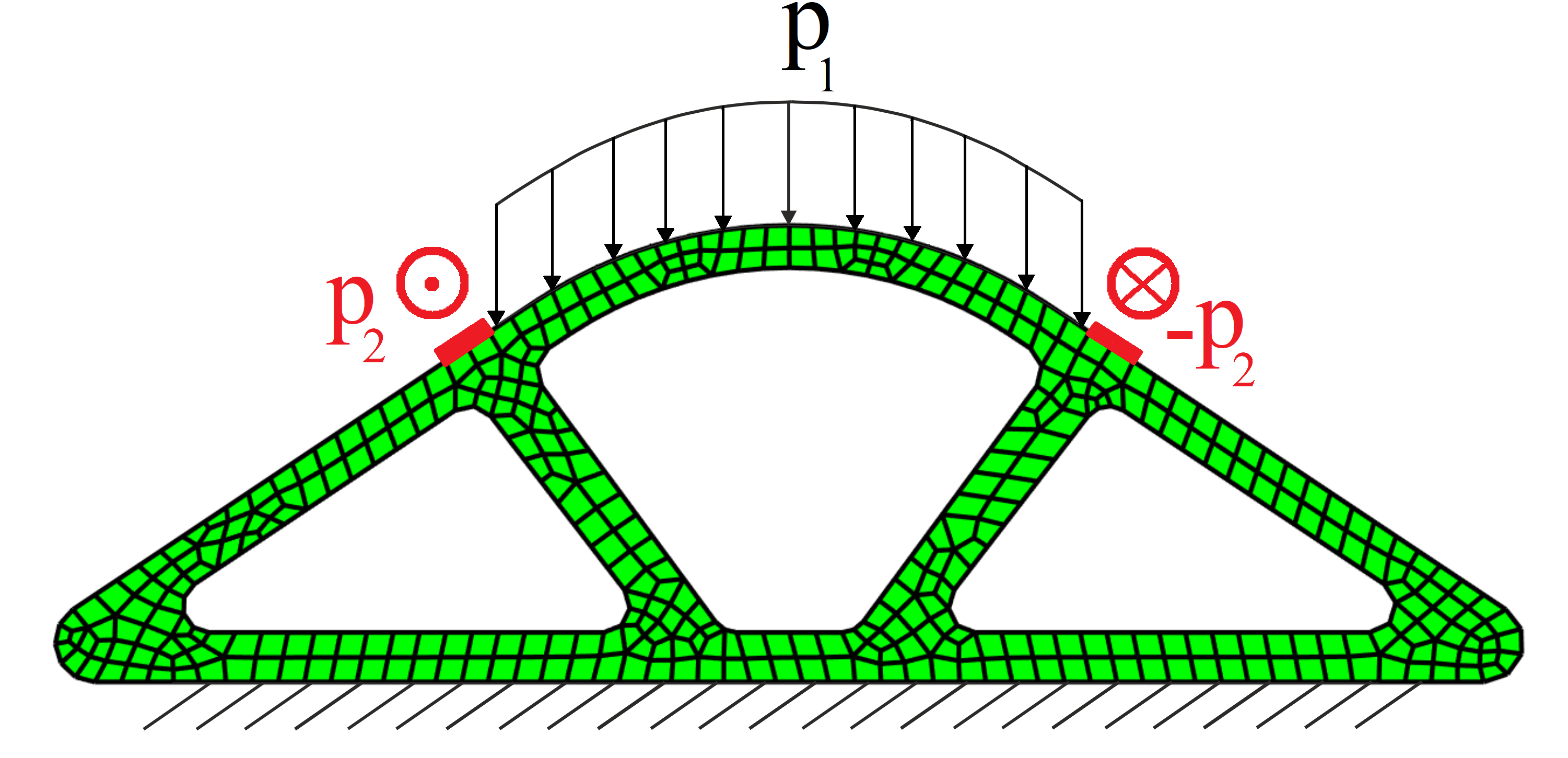}
\includegraphics[width=0.48\textwidth]{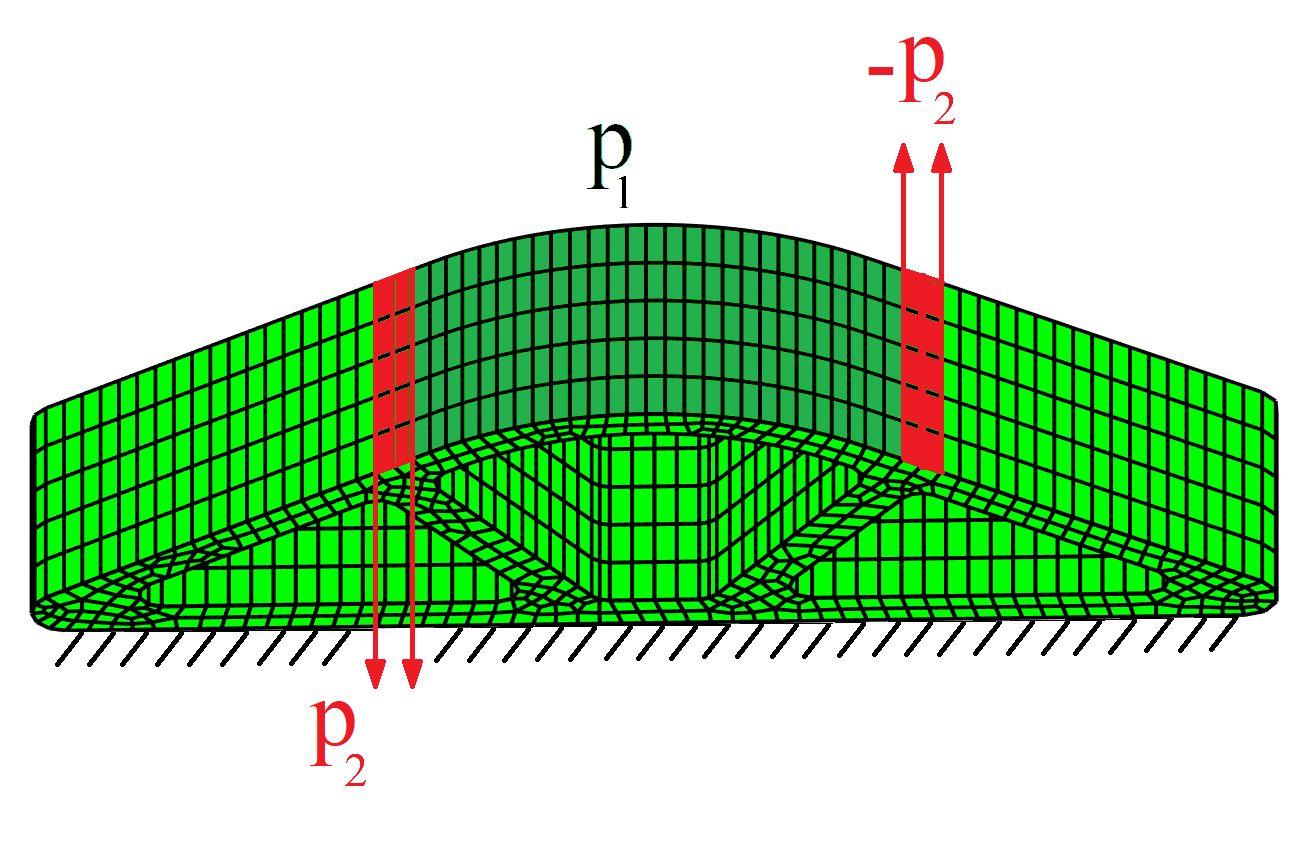}
\caption{Mesh and boundary conditions of the rubber sealing in Section~\ref{sec:exa3}. An out-of plane surface load in $z$-direction is applied (red); the elements effected by the load $p_1$ are in darker green.}\label{fig:examp3_bc}
\end{figure}

\paragraph{Macroscopic finite element model}
The geometry and boundary conditions are the same as before, see  Fig.~\ref{fig:_dichtung}, but the component has now a thickness of $5\,$mm. Additional to the surface load $p_1$  from atop we add two surface loads $p_2$ pointing into the transversal $z$-direction and thus twisting the component,
see Fig.~\ref{fig:examp3_bc}. 
We set $p_1=-7.5\,$kPa and $p_2=\pm 25\,$kPa. 

\begin{figure} 
\centering
\includegraphics[width=1\textwidth]{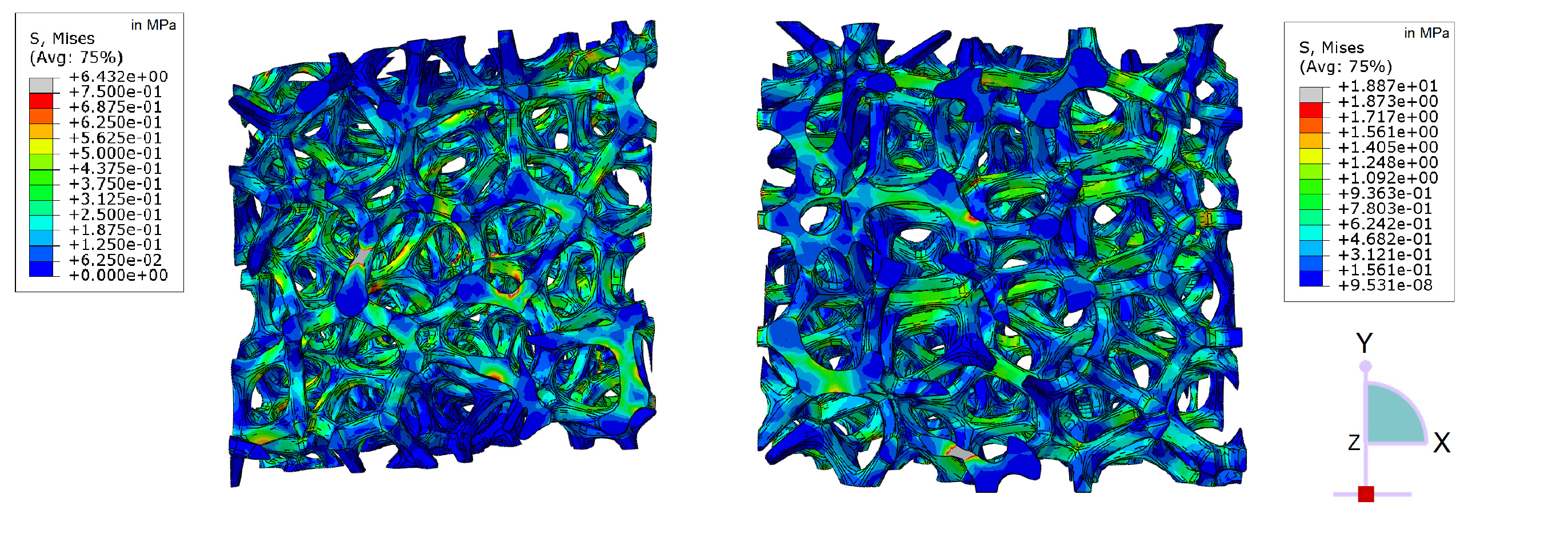}
\caption{Two of the six loading scenarios of case C: On the left side a shear deformation is displayed and on the right side the RVE that is introduced in section \ref{subsec:rve} is elongated along the x-axis.}\label{fig:rve_wurfel_simul}
\end{figure}

\paragraph{Generation of the material data set}
To mimic the polyurethane foam we use the RVE presented in Section \ref{subsec:rve} and compute it with the material data of 
Section~\ref{sec:exa3}. Under the loads of Fig.~\ref{fig:examp3_bc} we expect only moderate deformations and because of that we  consider a linear behavior. Due to the stochastic random sphere packing of the RVE, small deviations in the directions may still occur. Therefore, we chose the strategy outlined in Case C and conduct FEAs for the six unit loads of Fig.~\ref{rve_zustande}. The six  evaluated data tuples are listed in the appendix. We observe only  small deviations of around $5\%$  in the different directions; they are forwarded into the data set. %
Fig.~\ref{fig:rve_wurfel_simul} shows the RVE once sheared and once elongated in $x$ direction by $\epsilon=2\%$ and the relative low stresses. The material data set comprises stresses  between $-0.2\,$MPa and $0.2\,$MPa. The number of data tuples varies due to the multi-level approach and is listed for the different levels of refinement in Table~\ref{table:MLex3}.

\begin{table}
\centering
\ra{1.2}
\begin{tabular}{crc}\toprule
Level $l$  & data tuples in $\mathcal{D}_l$& data tuples in $\mathcal{S}_l$   \\ \hline
0 & 1.771.561 & 2.179    \\

1 & 1.588.491 & 8.889  \\

2 & 6.480.081 & 12.541  \\

3 & 9.142.389 & 14.152  \\

4 & 10.316.808 & 14.177  \\\hline
\bottomrule
\end{tabular}
\caption{Size of the material data set $\mathcal{D}$ and  number of data tuples describing the solution $\mathcal{S}$ in the multi level computation.  }
\label{table:MLex3}
\end{table}

\begin{figure} 
\centering
\includegraphics[width=0.98\textwidth]{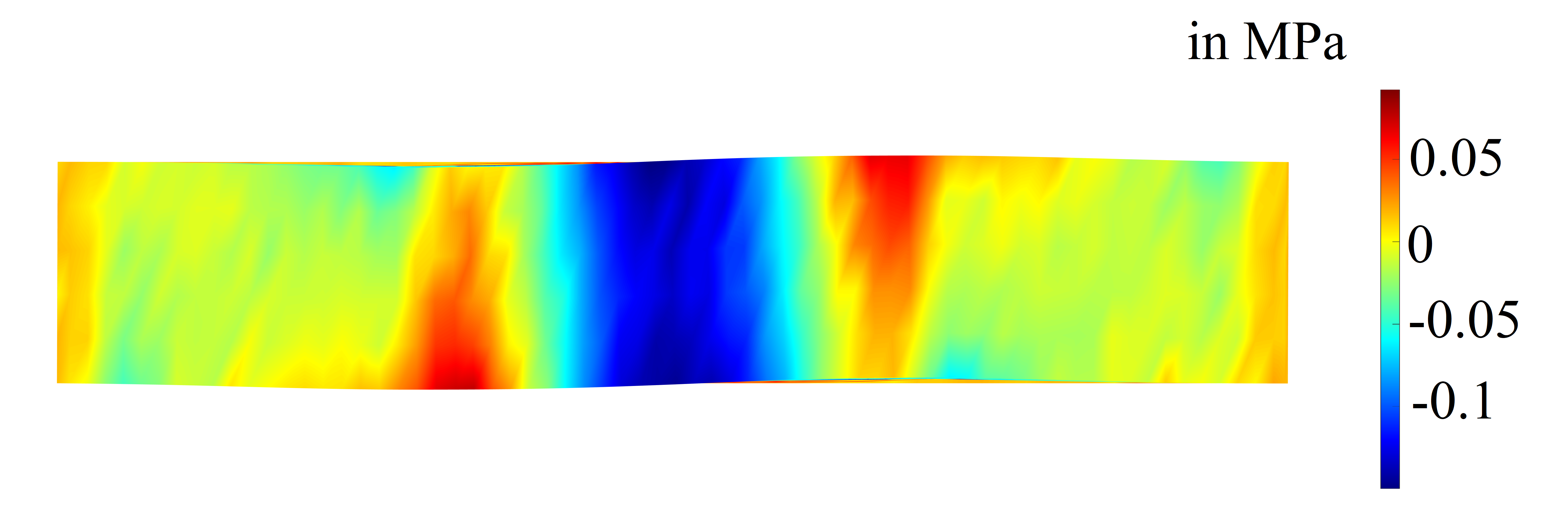}
\caption{View of the normal stress  $\sigma_{x}$ from above in the rubber sealing of the  DD-FEM (bottom). }\label{fig:examp3_11}
\end{figure}

\paragraph{Data-driven finite element analysis}
For the data-driven computation we used the multi-level method with 4 levels of refinement. The data refinement is stopped then because the increase of additional data points assigned to a material point is small in the last step. In Fig.~\ref{fig:examp3_11} the horizontal stress $\sigma_x$ is plotted for the DD-FEM. The displayed stress distribution shows that the results are satisfying for  $\sigma_x$.
 
\begin{figure} 
\centering
\includegraphics[width=0.98\textwidth]{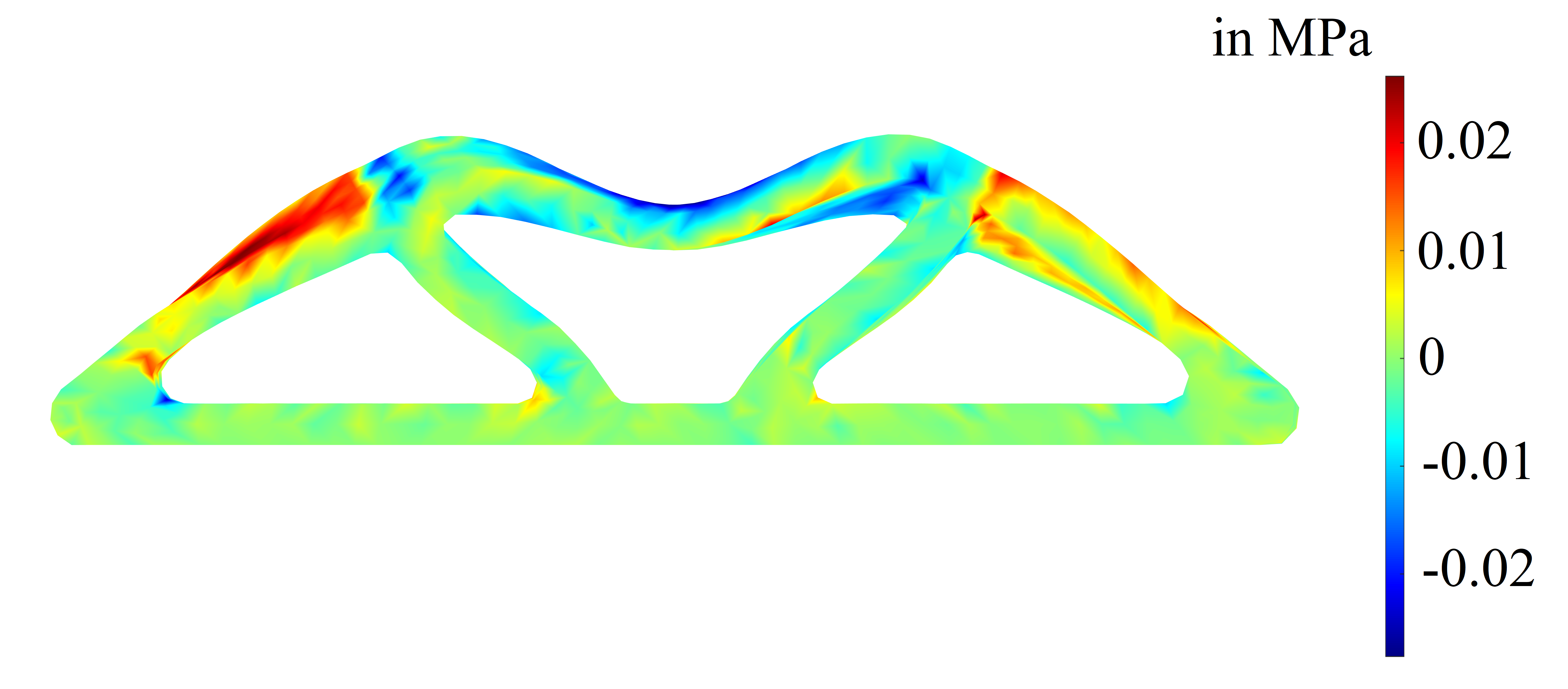}
\caption{Shear stress $\tau_{xz}$ in the rubber sealing for the  DD-FEM (level 4). The numerical generation allows to generate suitable data in all directions. }\label{fig:examp3_13}
\end{figure}

The corresponding shear stress $\tau_{xz}$  is displayed in Fig.~\ref{fig:examp3_13} in the deformed sealing. 
The displacement is magnified here by a factor of five so that the deformation of the component can nicely be seen. {The numerical generation of the data can be useful for smaller stress components. In  \cite{KorzeniowskiWeinberg2019} it is noted that for an equidistant data mesh smaller stress components could not be displayed adequately. The material density was derived by the maximal stress component and the number of data points. Since the data is generated numerically here, one would not be bound to a minimum recording resolution of an experimental setup and could still refine individual components here, which was not overused however.
Finally, the data-driven method manages to perform the macroscopic FEM simulation satisfactorily.} 

\section{Conclusion}\label{sec:summary}

In this paper, we propose a  strategy to provide the necessary material database for linear and non-linear data-driven finite element computations. Typically, such data are presumed to be available from experimental investigations but here we suggest a computational material testing instead. With representative  material volumes the micro-scale problem is investigated and the derived homogenized data give the input for the solution of the macro-scale problem.

Specifically we consider the DD-FEM of components made of open-cell foam. The representative volume elements map the 
foam's characteristics such as pore volume, pore distribution, pore and ligament geometry which are usually gained from CT scans. Their computation, in conjunction with simplifications made possible by the spatial material behavior, enable us to generate the material data input set.


We showed the capabilities of the methodology with three different examples. The first example is a study of how the data-driven method behaves with non-linear data and a  linear setting and the other way round. In the two rubber sealing examples we focus onto the material generation and simulation of the component. From our point of view, the DD-FEM and the data-based generation of the RVE can really excel the computation of such problems. With the methods presented here, it is possible to numerically calculate the properties of the foam, and the response of a component made of it, only with the help of CT scans. No further practical or experimental action is necessary. In particular, the DD method can help in the future design of new types of foam. If a new foam is to be designed, its effect on the final product can be determined in advance by varying individual target parameters in the creation of the RVE. By selecting the target parameters in the RVE, e.g. pore volume or pore size,  it is possible to determine which microscopic properties the foam has and how a component made of it would react to a given load. This whole process would be purely numerical and can save costs as well as time.

{Furthermore, the data-driven method can be used to insert material properties that were not inserted from a theoretical point of view, for example non-linear data in a linear simulation or anisotropy in an isotropic calculation.}
From our point of view, data-driven FEM is an interesting and developable methodology that should be given further attention in the future.


\section*{Acknowledgement}
We gratefully acknowledge the support of the Deutsche Forschungsgemeinschaft (DFG) under the project 
WE2525/10-2 as part of the Priority Program SPP1886 ``Polymorphic uncertainty modelling for the numerical design of structures'' (no. 273721697).

\section*{Declaration of interest}
The authors declare that there is no conflict of interest regarding the publication of this article.

\bibliography{lit_data,lit_rveschaum}

\setcounter{section}{0}
\renewcommand{\thesection}{\Alph{section}}%
\section{Appendix}\label{Appendix}
\subsection{Systems of equations for the \textbf{F},\textbf{P} formulation}
The linear finite element system of Eq.~\reff{DD:statik3D:FEMgleichungPF} and \reff{eq:Kraftvektor} reads:
\begin{align*}
    \M{K}^e_u {\Vhatu_e} = \M{f}^e_u  \ :\qquad&
    \M{K}_u^e=   \mu_0 \int_{\Omega^e} {\M{B}^e}^T \M{B}^e \td \Omega
    &\M{K}_u =   \bigcup_{\mathcal{E}} \M{K}^e \qquad\qquad\\\nonumber                                      &
    \M{f}^e_u = \mu_0 \int_{\Omega^e} {\M{B}^e}  {\boldsymbol{\epsilon}^*}\td \Omega
    &\M{f}_u =     \bigcup_{\mathcal{E}} \M{f}^e \qquad\qquad \
\\
    \M{K}^e_\lambda \Vhatv_e = \M{f}^e_\lambda  \ :\qquad&
    \M{K}_\lambda^e=   \mu_0 \int_{\Omega^e} {\M{B}^e}^T  \M{B}^e \td \Omega
    &\M{K}_\lambda =   \bigcup_{\mathcal{E}} \M{K}^e \qquad\qquad\\\nonumber                                      &
    \M{f}^e_\lambda = \M{f}^e -\int_{\Omega^e} {\M{B}^e}^T {\boldsymbol{\sigma}^*}\td \Omega
    &\M{f}_\lambda =     \bigcup_{\mathcal{E}} \M{f}^e \qquad\qquad \
\end{align*}

\subsection{Systems of equations for the \textbf{C},\textbf{S}  formulation}\label{appendixCS}
The non-linear finite element system of Eq.~\ref{eq:variSCsystemRfem} reads as residual equations  
\begin{align*}
    {\M R}_{u} &= \int_{\Omega_e} {\M{B}^e}^T\bigg(
    2\mu_0  {\M{F}^e}^T\Big({\M{F}^e}^T \M{F}^e -\M{C^*}\Big)    -   {\M{B}^e}^T \Vhatv \M{S}^*
    -
    \mu_0  {\M{B}^e}^T  \Vhatv {\M{B}^e}  \Vhatv {\M{F}^e}^T
    \bigg)     \td \Omega
    =0
    \\
    \M R_{\lambda} &= \int_{\Omega_e}
    \mu_0 {\M{B}^e}^T  {\M{F}^e}  {\M{B}^e}\Vhatv {\M{F}^e}^T
     +
    {\M{B}^e}^T  {\M{F}^e}  \M{S}^*
    \td \Omega
    - \M f^\text{ext} = 0
\end{align*}
 A corresponding Newton-Raphson iteration step has the form
\begin{align*}
       \begin{bmatrix}
         \Vhatu_{j+1} \\ \Vhatv_{j+1}
       \end{bmatrix} =
       \begin{bmatrix}
         \Vhatu_{j} \\ \Vhatv_{j}
       \end{bmatrix} +
       \begin{bmatrix}
         \Delta \Vhatu  \\ \Delta \Vhatv
       \end{bmatrix} \quad
       \text{  mit } \quad
        \begin{bmatrix}
         \M R_{u}(\Vhatu_{j},\Vhatv_{j}) \\ \M R_{\lambda}(\Vhatu_{j},\Vhatv_{j})
       \end{bmatrix} +
       \begin{bmatrix}
         \M K_{uu} & \M K_{u\lambda} \\
         \M K_{\lambda u} & \M K_{\lambda\lambda}
       \end{bmatrix}
       \begin{bmatrix}
         \Delta \Vhatu  \\ \Delta \Vhatv
       \end{bmatrix}     = \M 0
\end{align*}
where the $\M K$ terms  abbreviate the current tangent stiffness matrix. Their entries are calculated as
\begin{align*}
    \M K_{uu} = \frac{\partial \M R_u}{\partial \M{u}} =&\int_{\Omega_e} {\M{B}^e}^T\bigg( 2\mu_0{\M{B}^e}^T ({\M{F}^e}^T \M{F}^e -\M{C^*})\nonumber\\
    &+4 \mu_0{\M{F}^e}^T{\M{B}^e}^T \M{F}^e  -    \mu_0  {\M{B}^e}^T  \Vhatv {\M{B}^e}  \Vhatv {\M{B}^e}^T \bigg)\td \Omega   
\\
    \M K_{u\lambda} = \frac{\partial \M R_u}{\partial \boldsymbol{\lambda}} =&\int_{\Omega_e}-{\M{B}^e}^T{\M{B}^e}^T  \M{S}^*      -{\M{B}^e}^T\mu_0  {\M{B}^e}^T   {\M{B}^e}  \Vhatv {\M{F}^e}^T \nonumber\\
    &-{\M{B}^e}^T\mu_0  {\M{B}^e}^T  \Vhatv {\M{B}^e}   {\M{F}^e}^T\td \Omega
\\
    \M K_{\lambda u} = \frac{\partial \M R_\lambda}{\partial \M{u}} =&\int_{\Omega_e}\mu_0 {\M{B}^e}^T  {\M{B}^e}  {\M{B}^e}\Vhatv {\M{F}^e}^T+\mu_0 {\M{B}^e}^T  {\M{F}^e}  {\M{B}^e}\Vhatv {\M{B}^e}^T \nonumber\\
    &+{\M{B}^e}^T{\M{B}^e}^T  \M{S}^*  \td\Omega
\\
    \M K_{\lambda\lambda} = \frac{\partial \M R_\lambda}{\partial \boldsymbol{\lambda}} =& \int_{\Omega_e}
    \mu_0 {\M{B}^e}^T  {\M{F}^e}  {\M{B}^e} {\M{F}^e}^T\td \Omega
\end{align*}
In each time or load step and in each data iteration the iterative solution of the system is necessary.

\subsection{Data generation of the 3-D example}\label{appendix3}
For the three dimensional example in Section \ref{sec:exa3} the six deformations of case C are described by six homogenized strain values. Remind that in the linear regime $\T C \approx 2 \T \epsilon +1$ and we can also gain data of other strain and stress equivalents in this regime. Therefore, we adopt small strain notation which leads to the six strain tensors in Voigt notation
\begin{align}\label{eps_zustande}
\bar{\T \epsilon}^{(1)}=\begin{pmatrix}
\alpha \\ 0\\0\\0\\0\\0
\end{pmatrix},
\bar{\T \epsilon}^{(2)}=\begin{pmatrix}
0 \\ \alpha\\0\\0\\0\\0
\end{pmatrix},
\bar{\T \epsilon}^{(3)}=\begin{pmatrix}
0 \\ 0\\\alpha\\0\\0\\0
\end{pmatrix},
\bar{\T \epsilon}^{(4)}=\begin{pmatrix}
0 \\ 0\\0\\\alpha\\0\\0
\end{pmatrix},
\bar{\T \epsilon}^{(5)}=\begin{pmatrix}
0 \\ 0\\0\\0\\ \alpha\\0
\end{pmatrix},
\bar{\T \epsilon}^{(6)}=\begin{pmatrix}
0 \\ 0\\0\\0\\0\\\alpha
\end{pmatrix} \,.
\end{align}
with an arbitrary deformation defined by $\alpha\in\mathbb{R}$. The six computations with the introduced RVE lead with $\alpha=0.02$ to the six homogenized stresses $\sigma_i$ $i=1,\dots,6$:
\begin{align*}
\bar\sigma^{(1)}&=\begin{pmatrix} 9.13  &3.87  &  3.96 &  -0.00  &  0.00 &  -0.00 \end{pmatrix}^T \cdot 10^4 \text{ Pa}\\[0.10cm]
\bar\sigma^{(2)}&=\begin{pmatrix} 3.87 & 8.57 & 3.89 &  0.00 & 0.00 & -0.00\end{pmatrix}^T \cdot 10^4 \text{ Pa}\\[0.10cm]
\bar\sigma^{(3)}&=\begin{pmatrix} 3.96 & 3.89 & 8.89 & -0.00 & 0.00 & -0.00 \end{pmatrix}^T \cdot 10^4 \text{ Pa}\\[0.10cm]
\bar\sigma^{(4)}&=\begin{pmatrix} -0.00 &  0.00 &  0.00 & 1.94 & -0.00 & 0.00 \end{pmatrix}^T \cdot 10^4 \text{ Pa}\\[0.10cm]
\bar\sigma^{(5)}&=\begin{pmatrix}  0.00 & 0.00 & 0.00 &  0.00 &   1.96 &   0.00 \end{pmatrix}^T \cdot 10^4 \text{ Pa}\\[0.10cm]
\bar\sigma^{(6)}&=\begin{pmatrix} -0.00 & -0.00 & -0.00 & 0.00 & -0.00 &  1.99 \end{pmatrix}^T \cdot 10^4 \text{ Pa}\\[0.10cm]
\end{align*} 

\end{document}